# A global adaptive discretization of velocity space for discrete velocity methods in predictions of rarefied and multi-scale flows

Jianfeng Chen (陈健锋) ; Sha Liu (刘沙) ; Rui Zhang (张瑞) ; Congshan Zhuo (卓丛山) ; Yanguang Yang (杨彦广) ; Chengwen Zhong (钟诚文)

# A global adaptive discretization of velocity space for discrete velocity methods in predictions of rarefied and multi-scale flows

Jianfeng Chen (陈健锋),[1,a)] Sha Liu (刘沙),[1,2,3,b)] Rui Zhang (张瑞),[1] Congshan Zhuo (卓丛山),[1,2,3] Yanguang Yang (杨彦广),[4] and Chengwen Zhong (钟诚文)[1,2,3]

## AFFILIATIONS

[1]School of Aeronautics, Northwestern Polytechnical University, Xi'an, Shaanxi 710072, China
[2]Institute of Extreme Mechanics, Northwestern Polytechnical University, Xi'an, Shaanxi 710072, China
[3]National Key Laboratory of Aircraft Configuration Design, Northwestern Polytechnical University, Xi'an, Shaanxi 710072, China
[4]China Aerodynamics Research and Development Center, Mianyang, Sichuan 621000, China

[a)]Electronic mail: chenjf@mail.nwpu.edu.cn
[b)]Author to whom correspondence should be addressed: shaliu@nwpu.edu.cn

## ABSTRACT

By introducing a discrete velocity space (DVS), deterministic methods in gas-kinetic theory, such as the discrete velocity method (DVM) and unified methods, can accurately capture complex nonequilibrium distribution functions and describe rarefied flow behaviors. However, describing high-speed flows with conventional Cartesian DVS is prohibitively costly due to the large number of discrete velocity points. Therefore, to enable deterministic solvers to handle complex, rarefied, and multi-scale flows effectively, a novel adaptive velocity space (AVS) is proposed. First, a global velocity mesh is intentionally adopted instead of a local velocity mesh to maintain a high level of DVS parallelism and facilitate extension to implicit algorithms. The global AVS is robust as it avoids the instability of information transformation between different cell-local AVS. Second, a new strategy is developed for reconstructing the distribution function in the tree-structured AVS, which is a low-order reconstruction with forced macroscopic conservation. This low-order reconstruction facilitates the direct value assignment between farther and child nodes, avoiding the derivative calculation of the distribution function (which is sometimes discontinuous). Additionally, the integration error of the low-order reconstruction is fixed by the forced macroscopic conservation. With these two important improvements, the proposed global AVS is then integrated into general DVM frameworks, such as the unified gas-kinetic scheme. Finally, a number of numerical tests are carried out to validate the proposed method, including steady and unsteady multi-scale flows.

## I. INTRODUCTION

Multi-scale flows from earth's surface to outer space (or from macroscale to microscale) are common in scenarios such as near-space vehicles and microelectromechanical systems (MEMS), where multiple flow regimes (including continuum, slip, transitional, and free molecular ones) often coexist within a single flow field, leading to a complex dynamic process and challenging the physical modeling and numerical predictions. Different from the continuum fluid dynamics, in the case of multi-scale flows, the velocity distribution of molecules does not follow a normal distribution function, and should be recorded in detail. In earlier research on rarefied flows (in transitional and free molecular regimes), the nonequilibrium (abnormal) distributions were recorded in a statistical way or deterministic way. For statistical recording, model molecules are used, such as in the direct simulation Monte Carlo (DSMC) method.[1] For deterministic recording, the discrete velocity space (DVS) is adopted, and the distribution is recorded at discrete points in this DVS. The numerical methods using the DVS are known as the discrete velocity method (DVM) or the discrete ordinate velocity method (DOM).[2–5] Also based on the DVS, a class of unified methods, such as the unified gas-kinetic scheme (UGKS),[6,7] the discrete unified gas-kinetic scheme (DUGKS),[9,10] the gas-kinetic unified algorithm (GKUA),[4,8] and the improved discrete velocity method (IDVM),[11,12] have been proposed. With this detailed DVS and the aid of a time integral solution or characteristic line solution, these methods acquired the ability to solve the multi-scale flows in all regimes uniformly. At the present stage, unified methods have been successfully extended to other multi-scale physics, such as radiation of photons,[13,14] phonon heat transfer,[15,16] and plasma gas transfer.[17] After a decade of development, numerous numerical techniques have been devised and incorporated into these unified methods to enhance computational efficiency and reduce memory costs.[18–26] Due to the extra-DVS, the DVMs need huge computer memory and, therefore, huge computation time. To increase the computational efficiency, the implicit schemes designed for multi-scale computation are proposed, gaining an acceleration of 2–3 orders of magnitude. However, the fundamental cause of this substantial computational challenge remains the DVS, as it must both cover the range of molecular speeds and resolve the detailed distribution function. Taking a hypersonic flow with 10 times sonic speed (c), for example, the speed range of DVS is $(10 + 3)c$, while molecules often concentrate within a small region centered at $10c$ with a radius of $3c$ in velocity space and require a sufficient resolution (DVS points). For conventional uniform Cartesian DVS (CVS), about $10^6$ DVS points are required. Therefore, the unstructured DVS (UVS) that arranges the DVS points according to given distributions is proposed,[27–29] which is flexible and more efficient. While the arrangement of DVS points is empirical and case-dependent. Therefore, an adaptive velocity space (AVS) is in a strong demand. Aristov[30] proposed an AVS for the 1 D shock structure calculation. However, the approach is very specific to this test case and has never been extended. Chen *et al.*[18] and Kolobov and Arslanbekov[31] proposed to use different DVS for each point in the position space and every time. Soon after, Brull and Mieussens[32] proposed to use similar DVS while each DVS has its own bounds and steps that are evolved in time and space by using the local macroscopic velocity and temperature. Recently, Xiao *et al.*[33] replaced the discrete velocity points in the UGKS with a continuous velocity space in the region where the GKS solutions are valid.

For the present AVS methods, each physical cell has its own AVS.[18,31] The adaptation is according to the value and slope of the distribution function in velocity space. Specifically, more AVS points are arranged in the regions where the value or slope of the distribution function is large, and AVS points are removed in the regions where the value or slope of the distribution function is small enough. Since different cells (in physical space) have different arrangements of AVS points, when distribution information transfers between two cells (such as in the flux calculation), the distribution function in one cell should be interpolated in velocity space to match the AVS points in the other cell. Given that the distribution function is not always smooth (even not continuous) in velocity space, robust treatment should be adopted, and the conservation of mass, momentum, and energy should be forced. In this paper, these AVS methods are referred to as the local AVS, since the AVS is cell dependent.

At the present stage, both the UVS and AVS can reduce the velocity points by about an order of magnitude. While comparing with the UVS and the CVS, whose arrangement of velocity points is identical in all physical cells (referred to as the global DVS), implementing the implicit treatment and efficient DVS parallelism on local AVS is not easy, while their acceleration of 2–3 orders of magnitude is crucial for the huge computation works of DVMs. Therefore, in this work, a global AVS is proposed combining the advantages of both global DVS and local AVS (Fig. 1). Compared with the local AVS proposed by Chen *et al.*,[18] the present global AVS does not need to interpolate the flux between different DVS, making it perfectly compatible with the implicit algorithm. Additionally, the global AVS can employ velocity parallelism for efficient simulation without load balance challenges. Chen *et al.* employed a polynomial to redistribute the distribution function in the new AVS, with the polynomial parameters obtained by solving the conservation equations. In this work, a new strategy to simplify and stabilize the operations on the tree-structured AVS is proposed, utilizing an efficient forced conservation method previously proven on the UVS. Furthermore, the criteria of relative density and internal energy are combined with the multi-zone strategy, which refines the core region of the DVS, to guide the adaptation process. To test the validity of the present AVS, it is implemented on a famous multi-scale DVM, the UGKS method.

The rest of the paper is organized as follows: In Sec. II, the details of the global adaptive UGKS are discussed, including the UGKS with simplified multi-scale numerical flux, the adaptive criteria, the adaptive process, the reconstruction of distribution function, the conservation correction, and the automatic domain setting. In Sec. III, a series of multi-scale steady and unsteady flow simulations are conducted to verify the proposed global AVS, including the steady ones and unsteady ones. Finally, the conclusion is made in Sec. IV.

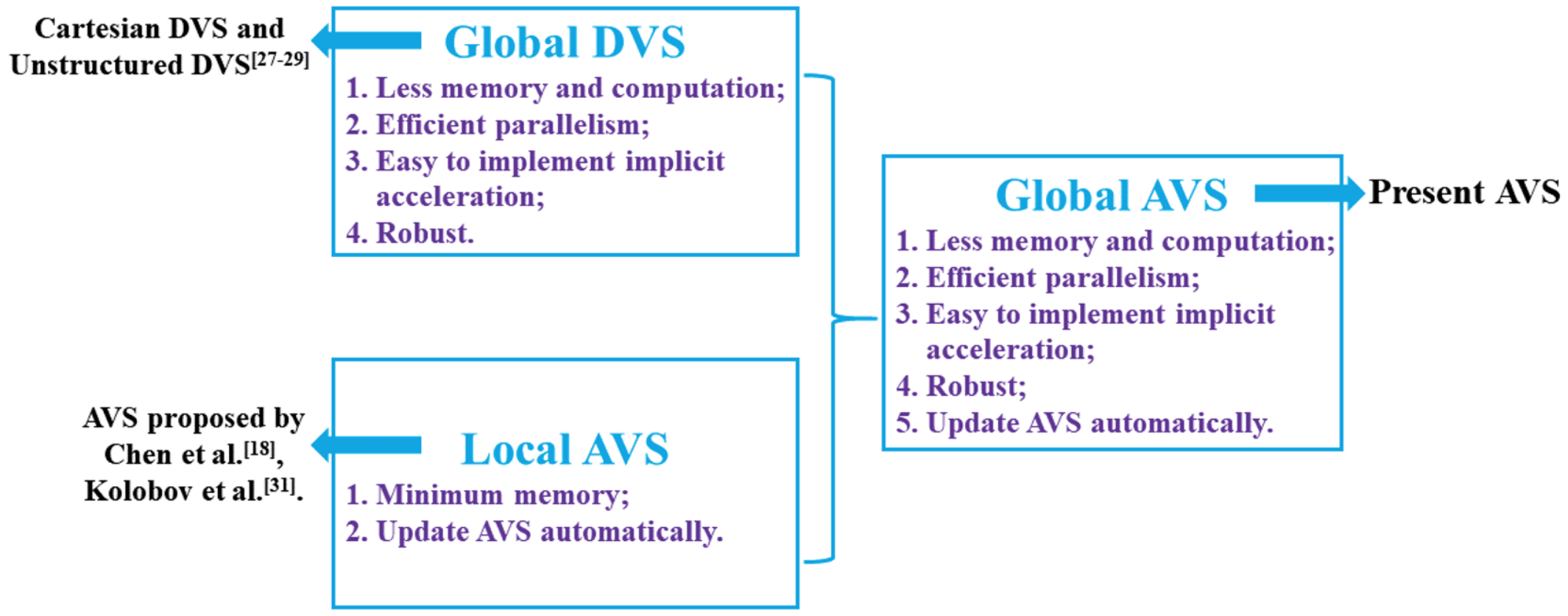

**FIG. 1.** A global AVS is proposed combining the advantages of both global DVS and local AVS.

## II. GLOBAL ADAPTIVE DISCRETE VELOCITY SPACE

The present global AVS is proposed for general discrete velocity methods, and in this section, it is implemented in a UGKS method with simplified multi-scale flux, whose details can be found in the literature.[29,34] Then, the main body of this section focuses on the details of global AVS. For completeness, the present DVS suits the implicit scheme and diatomic gases.

### A. Unified gas-kinetic scheme with simplified multi-scale numerical flux

The UGKS adopts the gas-kinetic relaxation model equation in the following form:

$$\frac{\partial f}{\partial t} + \xi \cdot \nabla f = \Omega \equiv \frac{g-f}{\tau}, \tag{1}$$

where $f = f(\boldsymbol{x}, \boldsymbol{\xi}, \boldsymbol{\eta}, e, t)$ is the distribution function for particles moving in the D-dimensional physical space with a velocity of $\boldsymbol{\xi} = (\xi_1, \ldots, \xi_D)$ at position $\boldsymbol{x} = (x_1, \ldots, x_D)$ and time $t$. Here, $\boldsymbol{\eta} = (\xi_{D+1}, \ldots, \xi_3)$ is the dummy velocity consisting of the rest components of the particle velocity in three-dimensional (3 D) space, $e$ represents molecular rotational energy. $\Omega$ is the collision operator. $\tau$ is the relaxation time relating to the dynamic viscosity $\mu$ and pressure $p$ with $\tau = \mu/p$. $g$ is the equilibrium distribution function, such as the Maxwellian equilibrium distribution $g^{eq}$, the Shakhov[35] equilibrium distribution $g^S$, and the Rykov[36] equilibrium distribution $g^R$,

$$g^{eq} = \frac{\rho}{(2\pi RT)^{(3+K)/2}} \exp\left(-\frac{c^2+\eta^2+e^2}{2RT}\right), \tag{2}$$

$$g^S = g^{eq}\left[1 + (1-\mathrm{Pr})\frac{\boldsymbol{c}\cdot\boldsymbol{q}}{5pRT}\left(\frac{c^2+\eta^2}{RT} - 5\right)\right], \tag{3}$$

$$g^R = \left(1 - \frac{1}{Z_{rot}}\right) g^{tr} + \frac{1}{Z_{rot}} g^{rot}, \tag{4}$$

where $\rho$ is the density, $R$ is the gas constant, $T$ is the temperature, $Pr$ is the Prandtl number, $\boldsymbol{c} = \boldsymbol{\xi} - \boldsymbol{U}$ is the peculiar velocity with $\boldsymbol{U}$ being the macroscopic flow velocity, $\boldsymbol{q}$ is the heat flux, and $Z_{rot}$ is the rotational collision number. The distribution functions $g^{tr}$ and $g^{rot}$ in Eq. (4) are given by

$$\begin{aligned} g^{tr} &= n\left(\frac{1}{2\pi RT_{tr}}\right)^{\frac{3}{2}} \exp^{-\frac{c^2+\eta^2}{2RT_{tr}}} \frac{1}{mRT_{rot}} \exp^{-\frac{e}{mRT_{rot}}} \\ &\times \left[1 + \frac{\boldsymbol{c}\cdot\boldsymbol{q}_{tr}}{15RT_{tr}p_{tr}}\left(\frac{c^2+\eta^2}{RT_{tr}} - 5\right) \right. \\ &\left. + (1-\delta)\frac{\boldsymbol{c}\cdot\boldsymbol{q}_{rot}}{RT_{tr}p_{rot}}\left(\frac{e}{mRT_{rot}} - 1\right)\right], \end{aligned} \tag{5}$$

$$\begin{aligned} g^{rot} &= n\left(\frac{1}{2\pi RT}\right)^{\frac{3}{2}} \exp^{-\frac{c^2+\eta^2}{2RT}} \frac{1}{mRT} \exp^{-\frac{e}{mRT}} \\ &\times \left[1 + \omega_0\frac{\boldsymbol{c}\cdot\boldsymbol{q}_{tr}}{15RTp}\left(\frac{c^2+\eta^2}{RT} - 5\right) \right. \\ &\left. + \omega_1(1-\delta)\frac{\boldsymbol{c}\cdot\boldsymbol{q}_{rot}}{RTp}\left(\frac{e}{mRT} - 1\right)\right], \end{aligned} \tag{6}$$

where $n$ is the molecular number density, $m$ is the molecular mass, $T_{tr}$ and $T_{rot}$ are the translational and rotational temperature, respectively, $p_{tr}$ and $p_{rot}$ are the pressures corresponding to $T_{tr}$ and $T_{rot}$, respectively, $\boldsymbol{q}_{tr}$ and $\boldsymbol{q}_{rot}$ are the translational and rotational heat flux, respectively. The other coefficients are $\delta = 1/1.55$, $\omega_0 = 0.2354$, and $\omega_1 = 0.3049$ for nitrogen.[34] In this work, the Shakhov equilibrium distribution $g^S$ and the Rykov equilibrium distribution $g^R$ are adopted for the simulations of monatomic and diatomic gas flows, respectively.

Integrating Eq. (1) over control volume (cell) $j$ from time $t_n$ to $t_{n+1}$, the discrete governing equation can be written as

$$f_j^{n+1} - f_j^n + \frac{\Delta t}{|V_j|} F_j^{n+1/2} = \frac{\Delta t}{2}\left(\Omega_j^{n+1} + \Omega_j^n\right), \tag{7}$$

where $|V_j|$ is the volume of cell $j$, $\Delta t = t_{n+1} - t_n$ is the time step, and $F_j^{n+1/2}$ is the microscopic flux,

$$F_j^{n+1/2} = \sum_k \xi \cdot \boldsymbol{A}_j^k f\left(\boldsymbol{x}_j^k, \xi, t_{n+1/2}\right), \tag{8}$$

where $\boldsymbol{A}_j^k$ is the outward normal vector of the $k$th face of cell $j$ with an area of $|A_j^k|$, and $\boldsymbol{x}_j^k$ is the center of this face. In this study, a simplified multi-scale flux is employed. Integrating Eq. (1) along the characteristic line (in the direction of particle velocity) from $t_n$ to $t_{n+1/2}$, the interface distribution function $f(\boldsymbol{x}_j^k, \xi, t_{n+1/2})$ can be expressed as

$$f\left(\boldsymbol{x}_j^k, \xi, t_{n+1/2}\right) = \frac{2\tau^{n+1/2}}{2\tau^{n+1/2}+\Delta t} f\left(\boldsymbol{x}_j^k - \xi\frac{\Delta t}{2}, \xi, t_n\right) + \frac{\Delta t}{2\tau^{n+1/2}+\Delta t} g\left(\boldsymbol{x}_j^k, \xi, t_{n+1/2}\right), \quad (9)$$

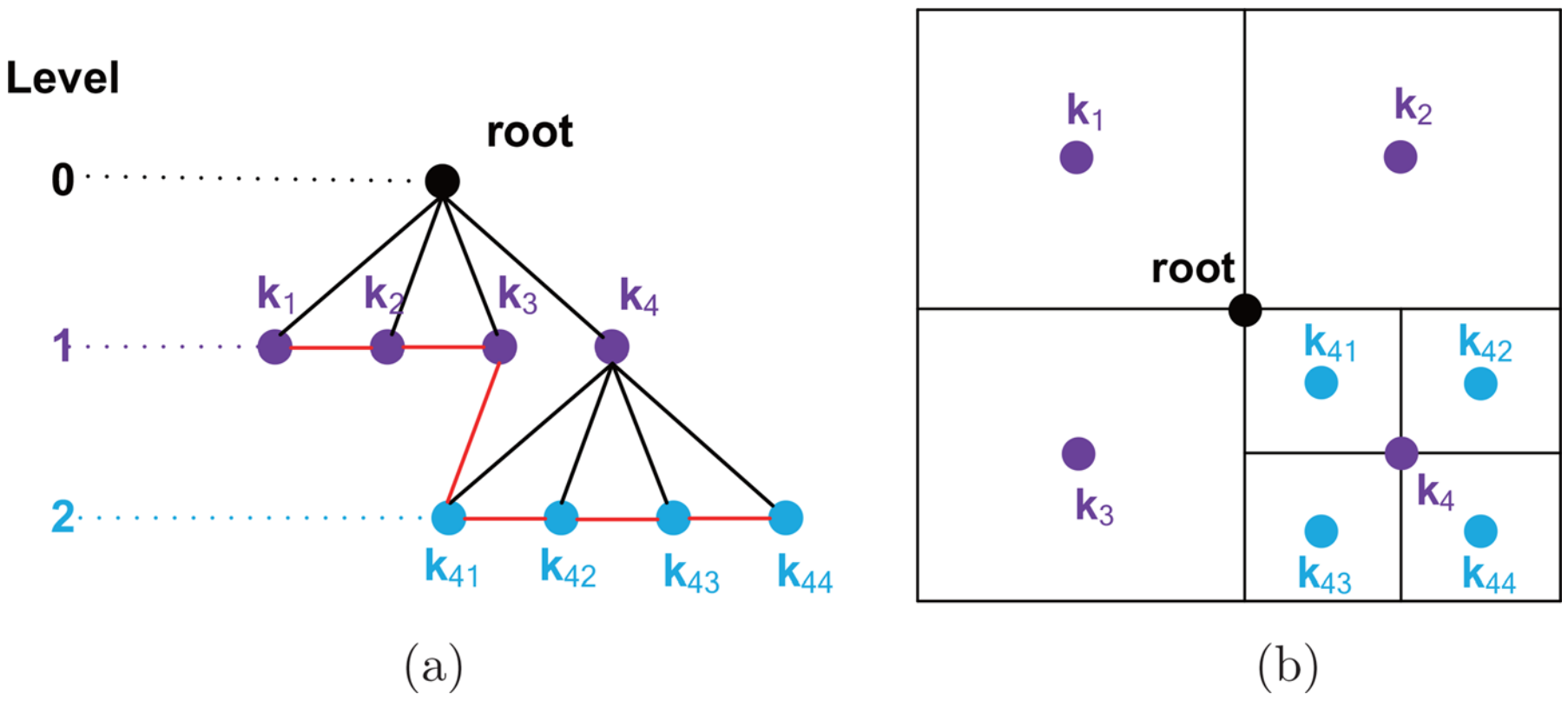

FIG. 2. Illustration of (a) quadtree data structure and (b) the corresponding velocity mesh.

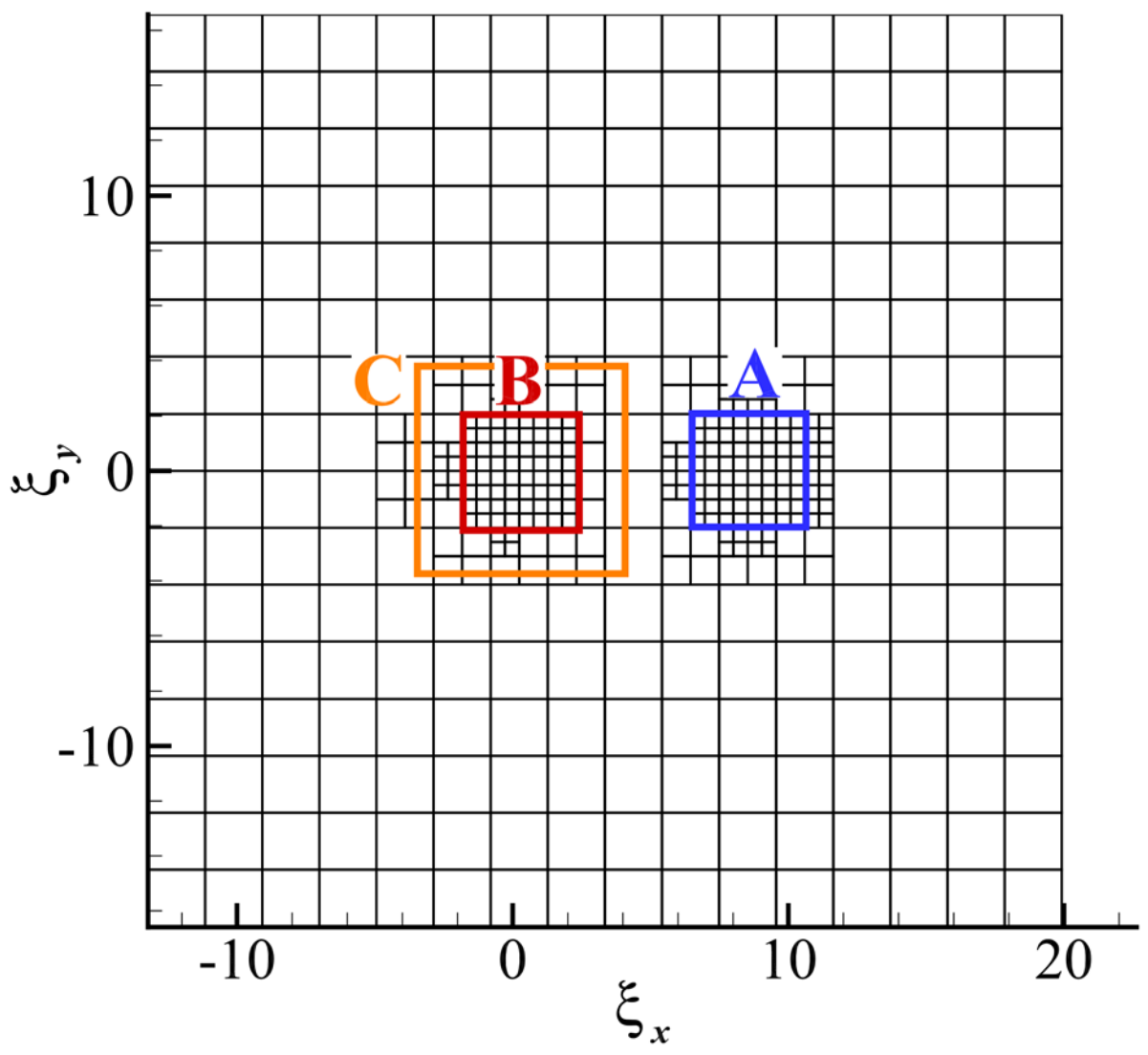

FIG. 3. The diagram of the global AVS.

TABLE I. The sub-zones of global AVS for local high resolution region.

| Zone | Center | Radius | Level | Degree of concentration |
|---|---|---|---|---|
| A | $\boldsymbol{U}_\infty$ | $3\sqrt{RT_\infty}$ | $L_{max}$ | Highest |
| B | $\boldsymbol{0}$ | $3\sqrt{RT_w}$ | $L_{max}$ or $L_{max}-1$ | High |
| C | $\boldsymbol{0}$ | $5\sqrt{RT_w}$ | $max(L_{min}, L_B - 1)$ | Sub-high |

where the free transport distribution (representing the microscopic mechanism) and equilibrium distribution (representing the macroscopic mechanism) are coupled rationally, leading to the multi-scale property of the present UGKS.

To run the microscopic evolution for getting $f$, the equilibrium state $g$ is needed, which is depended on the macroscopic variables $\boldsymbol{W} = (\rho, \rho\boldsymbol{U}, \rho E, \rho E_{rot})^T$, where $\rho$, $\rho\boldsymbol{U}$, $\rho E$, and $\rho E_{rot}$ are the density, momentum, total energy, and rotational energy, respectively. Therefore, take the moments of Eq. (7), the macroscopic evolution equation can be found as follows:

$$\boldsymbol{W}_j^{n+1} = \boldsymbol{W}_j^n - \frac{\Delta t}{|V_j|}\int \psi F_j^{n+1/2} d\xi + \frac{1}{2}\left(\boldsymbol{S}_j^{n+1} + \boldsymbol{S}_j^n\right), \quad (10)$$

where $\int \psi F_j^{n+1/2} d\xi$ is the macroscopic flux, $\psi$ is the collision invariant, and $\boldsymbol{S}$ is the source term.[37] Once the macroscopic variables are updated, the implicit equation Eq. (7) can be explicitly written as

$$f_j^{n+1} = \left(1 + \frac{\Delta t}{2\tau_j^{n+1}}\right)^{-1} \left[f_j^n - \frac{\Delta t}{|V_j|} F_j^{n+1/2} + \frac{\Delta t}{2}\left(\frac{g_j^{n+1}}{\tau_j^{n+1}} + \frac{g_j^n - f_j^n}{\tau_j^n}\right)\right]. \tag{11}$$

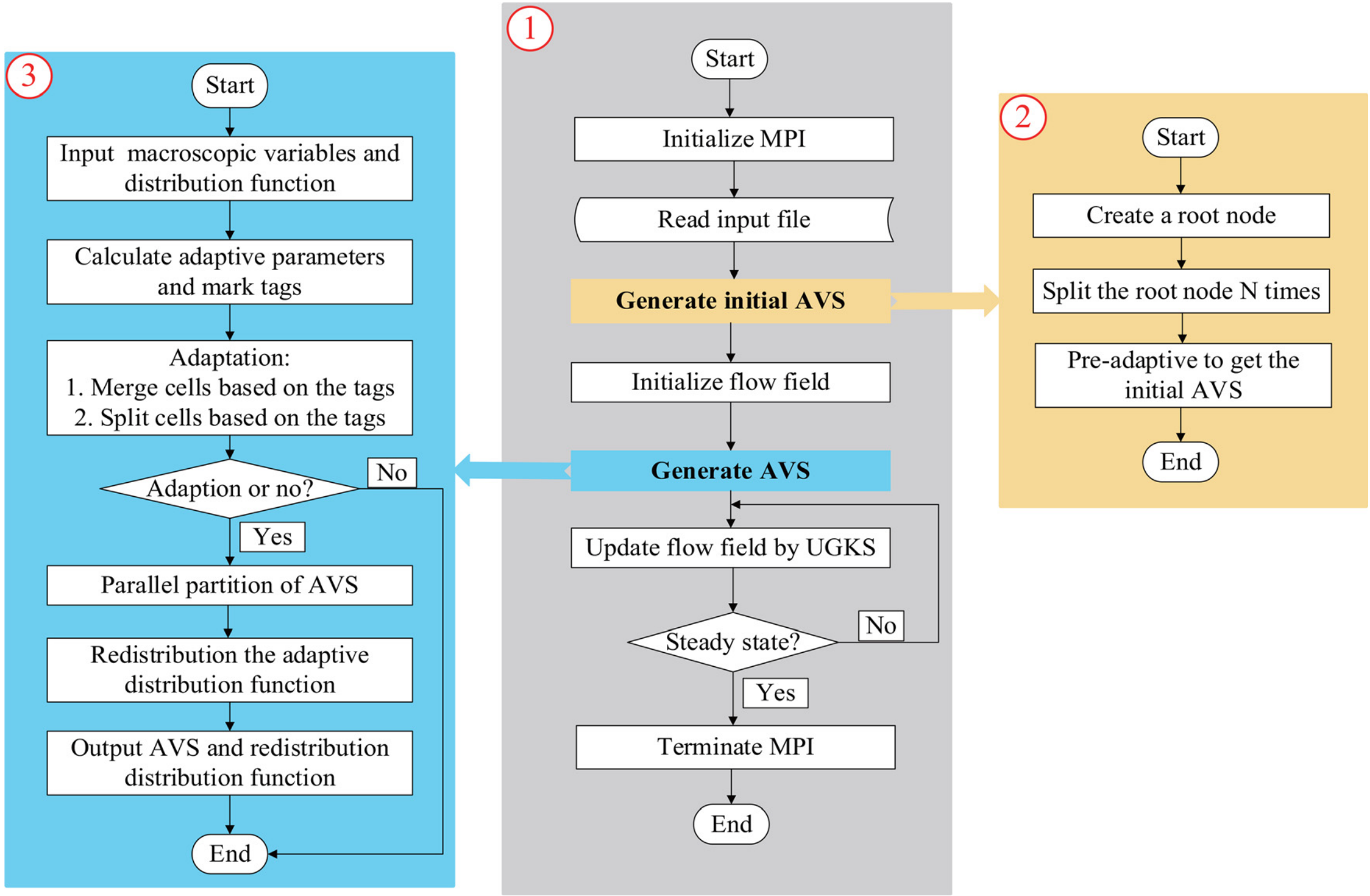

FIG. 4. The flow chart of global adaptive UGKS (The gray region represents the main trunk, while the orange and blue regions depict branches for generating the initial AVS and matching the AVS with the flow field, respectively.).

TABLE II. AVS of the hypersonic flow over a cylinder.

| Ma | $C_1$ | $L_{max}$ | Element |
|---|---|---|---|
| 2 | 5.0 | 4 | 136 |
| 4 | 2.0 | 5 | 268 |
| 5 | 1.3 | 5 | 232 |
| 6 | 1.0 | 6 | 670 |
| 8 | 0.5 | 6 | 526 |
| 10 | 0.5 | 6 | 472 |
| 12 | 0.5 | 7 | 796 |
| 14 | 0.3 | 7 | 748 |
| 16 | 0.3 | 7 | 658 |
| 18 | 0.3 | 7 | 610 |
| 20 | 0.2 | 7 | 676 |
| 22 | 0.1 | 7 | 856 |
| 24 | 0.1 | 8 | 1348 |
| 26 | 0.1 | 8 | 1360 |
| 28 | 0.1 | 8 | 1270 |
| 30 | 0.1 | 8 | 1174 |

Therefore, in UGKS, the macroscopic variables and distribution functions are updated sequentially by Eqs. (10) and (11), respectively.

## B. Details of global adaptive discrete velocity space

In this paper, for constructing the global AVS, the adaptive tree-data-based Cartesian mesh is adopted. The global AVS consists of three important aspects: (1) the adaptive criteria and adaptive process, (2) the forced macroscopic conservation, and (3) the automatic domain setting.

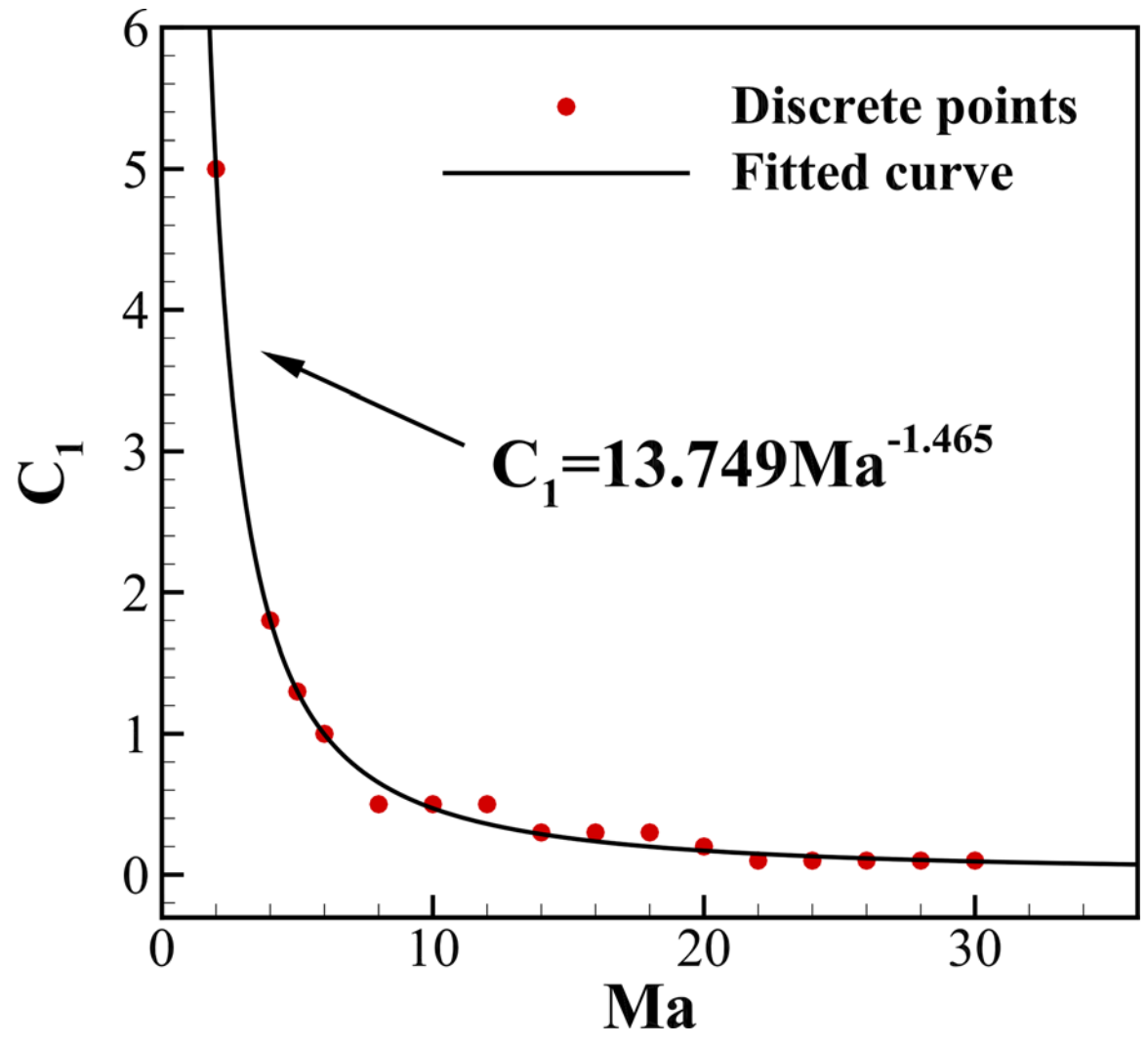

FIG. 5. The fitted curve of the parameter $C_1$ to the Mach number.

### 1. Adaptive criteria and adaptive process

Generally, the adaptation procedure involves a merging step and a splitting step. Therefore, adaptive criteria are necessary to determine whether a velocity element is going to merge, split, or remain unchanged before adaptation. In this work, the following density criterion and the internal energy criterion, referred to as the $M_1$ and $M_2$ criteria, respectively, are employed,[18] and the variables used are as follows:

$$M_{1,j,k} = A_k f_{j,k}/\rho_j, \tag{12}$$

$$M_{2,j,k} = \frac{1}{2} A_k f_{j,k} (\boldsymbol{\xi}_k - \boldsymbol{U}_j)^2 \Big/ \left[(\rho E)_j - \frac{1}{2}\rho_j \boldsymbol{U}_j^2\right], \tag{13}$$

where $j$ and $k$ denote the indices of the physical cell and velocity element (a DVS point is the center of the corresponding velocity element), respectively. $A_k$ is the area (2D) or volume (3D) of the $k$th

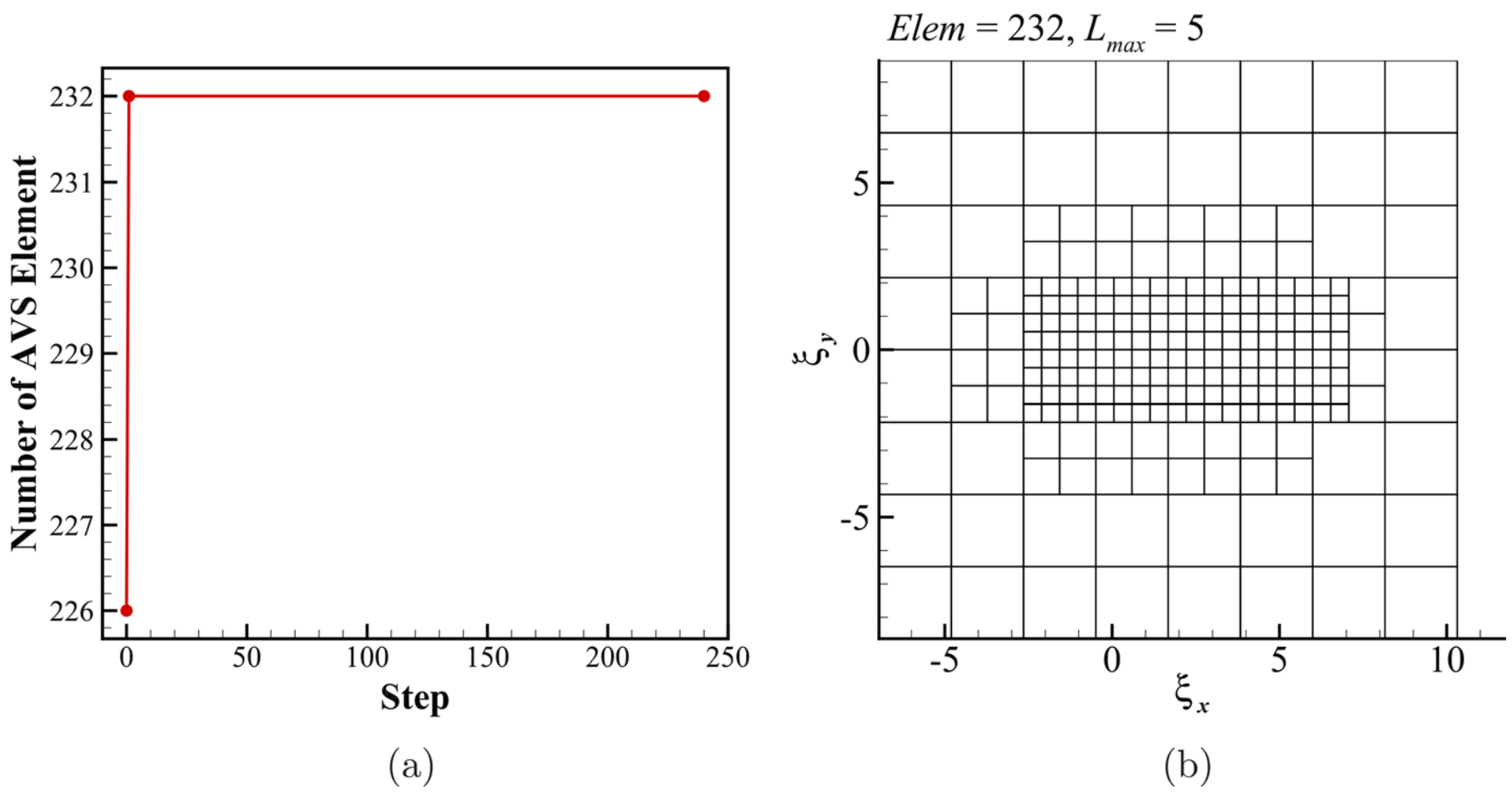

FIG. 6. AVS for the cylinder simulation at Ma = 5 and Kn = 1.0. (a) Evolution of the number of AVS elements and (b) the final AVS.

**TABLE III.** Computational efficiency of the cylinder simulation ($Ma = 5$).

| ... | CVS | AVS | Ratio |
|---|---|---|---|
| Elements | 7921 | 232 | 34.1 |
| Steps | 230 | 240 | 1.0 |
| Time (s) | 2223.0 | 99.4 | 22.4 |

velocity element. It can be observed that $M_1$ and $M_2$ represent the percentage of density and internal energy within the $k$th velocity element, respectively. If $M_1$ and $M_2$ are too large or too small, the velocity element should be split or merged. The present AVS is a global one that is the same in all physical cells. Therefore, the principle of the adaptation is: for a given velocity element, (1) if a $M_k$ with value larger than the splitting threshold is found in all physical cells, this velocity element splits; (2) if the maximum value of $M_k$ among all physical is still less than the merging threshold, this velocity element merges. This principle is aimed to guarantee the precision of AVS. Therefore, the maximum value of $M_k$ in the entire flow field is defined as follows:

$$M_k = \max(M_{1,j,k}, M_{2,j,k}), \quad j = 1, 2, \ldots, N_p, \tag{14}$$

where $N_p$ is the number of physical cells in the flow domain. Notice that the maximum of $M_1$ and $M_2$ is also chosen following the adaptive principle that makes splitting easy and merging hard for precision purposes. Given the splitting threshold $C_1$ and merging threshold $C_2$, the adaptive rules can be written as

(a) If $M_k > C_1$, the $k$th velocity element undergoes a splitting process.
(b) If $M_k < C_2$, the $k$th velocity element undergoes a merging process.
(c) If $C_2 \le M_k \le C_1$, the $k$th velocity element remains unchanged.

Since $C_1$ and $C_2$ govern the splitting and merging processes, respectively, their values are important. Being the same as the threshold of adaptive Cartesian mesh for physical space, the threshold for AVS is semiempirical. $C_1$ is chosen such that a series of standard test cases yield satisfactory results, and the tree height (or level $L$) is limited. The collaboration between $C_1$ and $L$ ensures that the data structure is not too deep around narrow distributions, while the distribution function points are sufficient in DVS regions with small value distributions. However, $C_2$ can be chosen more rationally by using the normal (Maxwellian) distribution as a reference distribution, and its value is related to $C_1$ via

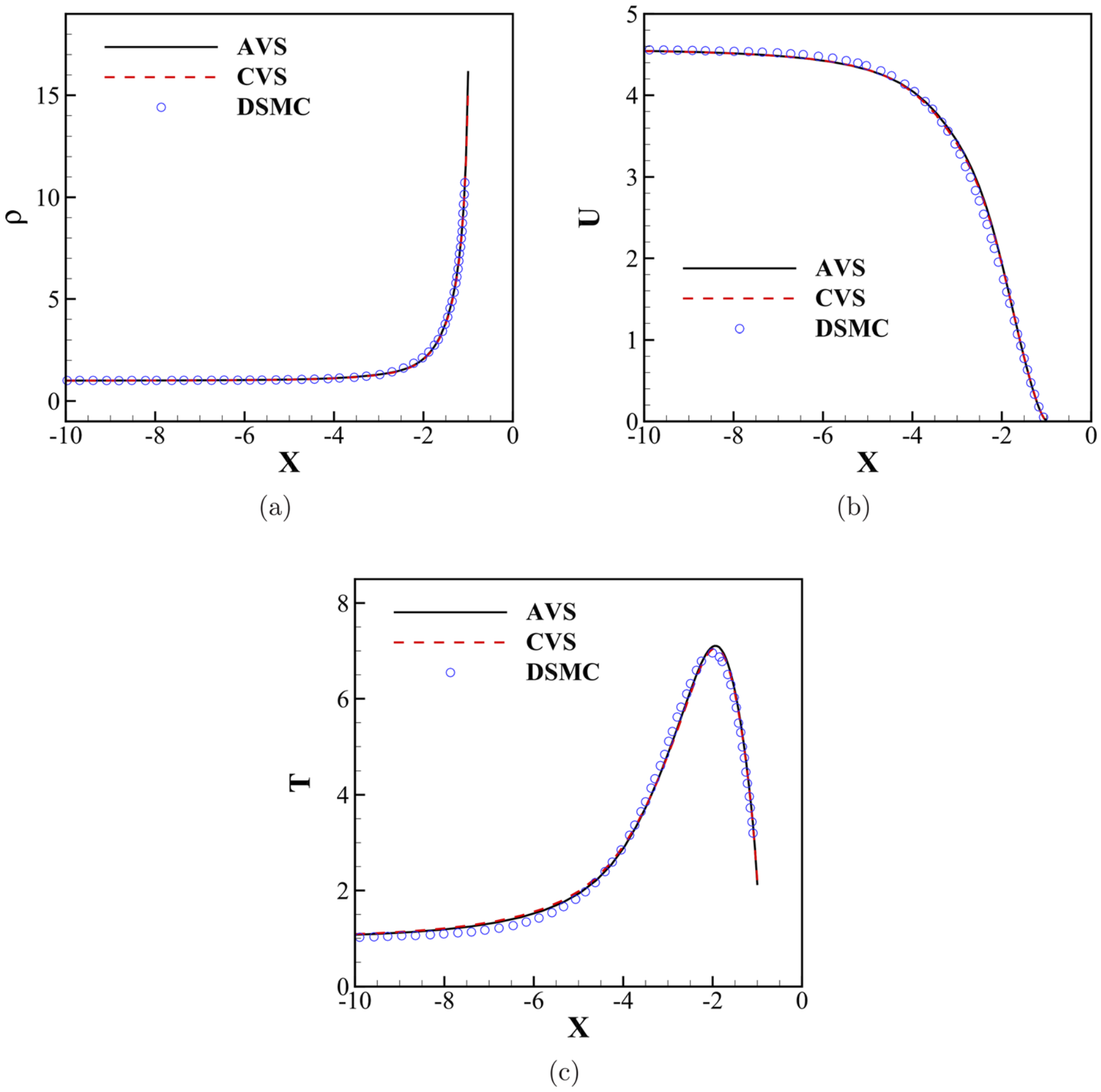

**FIG. 7.** Physical variables along the stagnation line in front of the cylinder at Ma = 5 and Kn = 1.0. (a) Density, (b) horizontal velocity, and (c) temperature.

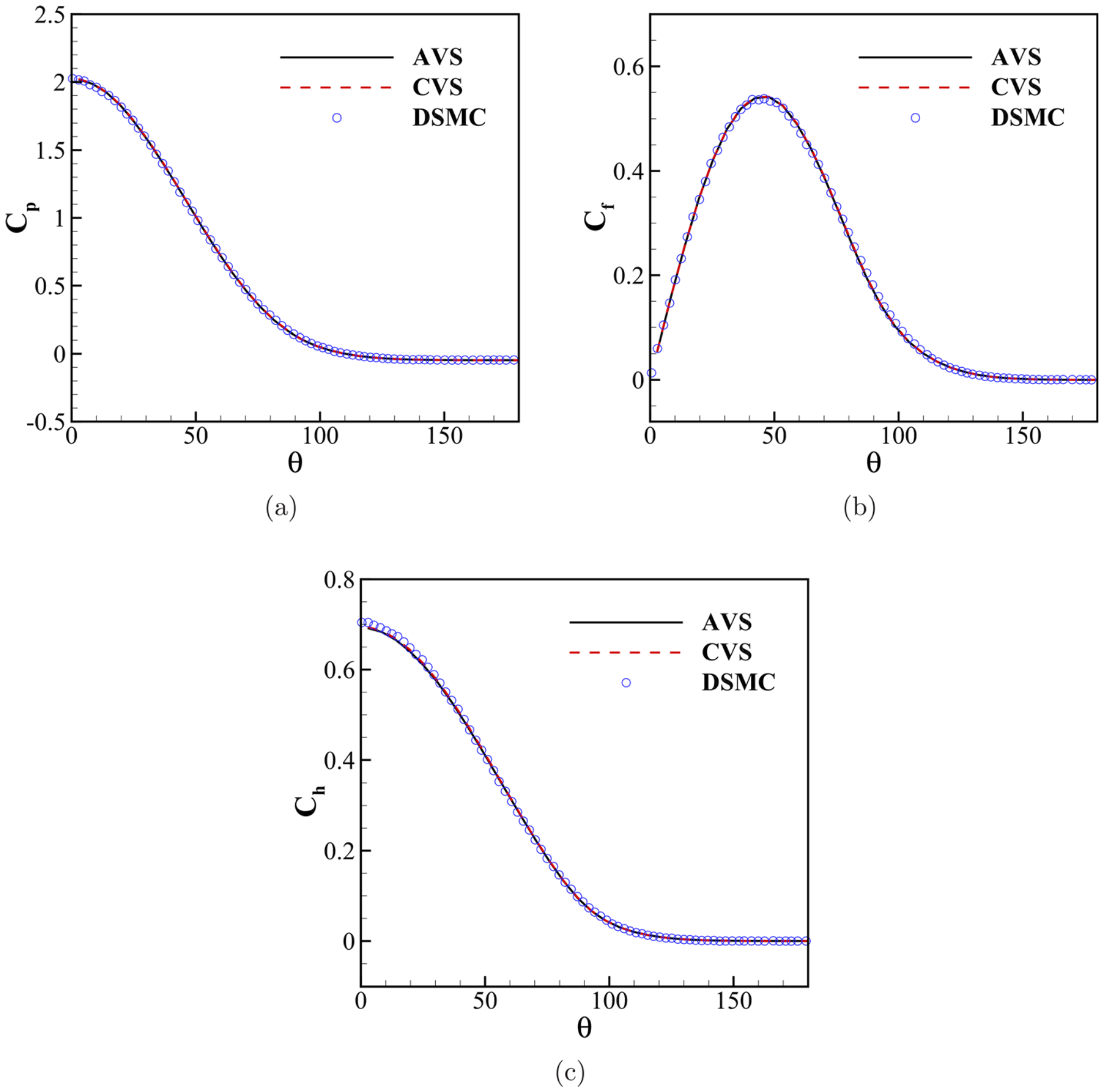

FIG. 8. Physical variables on the surface of cylinder at Ma = 5 and Kn = 1.0. (a) Pressure coefficient, (b) skin friction coefficient, and (c) heat transfer coefficient.

$$C_2 = k_c C_1, \tag{15}$$

where $k_c$ should be less than 1. According to the properties of the normal distribution, the probability within the $3\sigma$ radius and $4\sigma$ radius around the mean is 99.73% and 99.99%, respectively. Therefore, the $3\sigma$ radius and $4\sigma$ radius can be chosen as critical values. Considering the maximum value of the standard normal distribution $f_{max} = 0.3989$ and the critical value $f_{4\sigma} = 1.3383 \times 10^{-4}$, parameter $k_c$ can be calculated as follows:

$$k_c = \frac{f_{4\sigma}}{f_{max}} = 3.3546 \times 10^{-4}. \tag{16}$$

In this study, quadtree and octree data structures are employed to represent the 2 D and 3 D AVS, respectively. For the sake of simplicity,

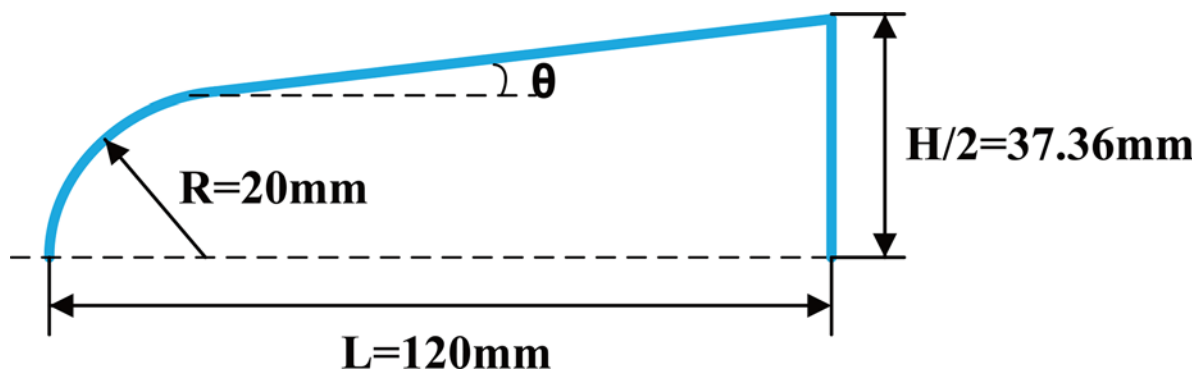

FIG. 9. Geometry of the blunt wedge (half-model).

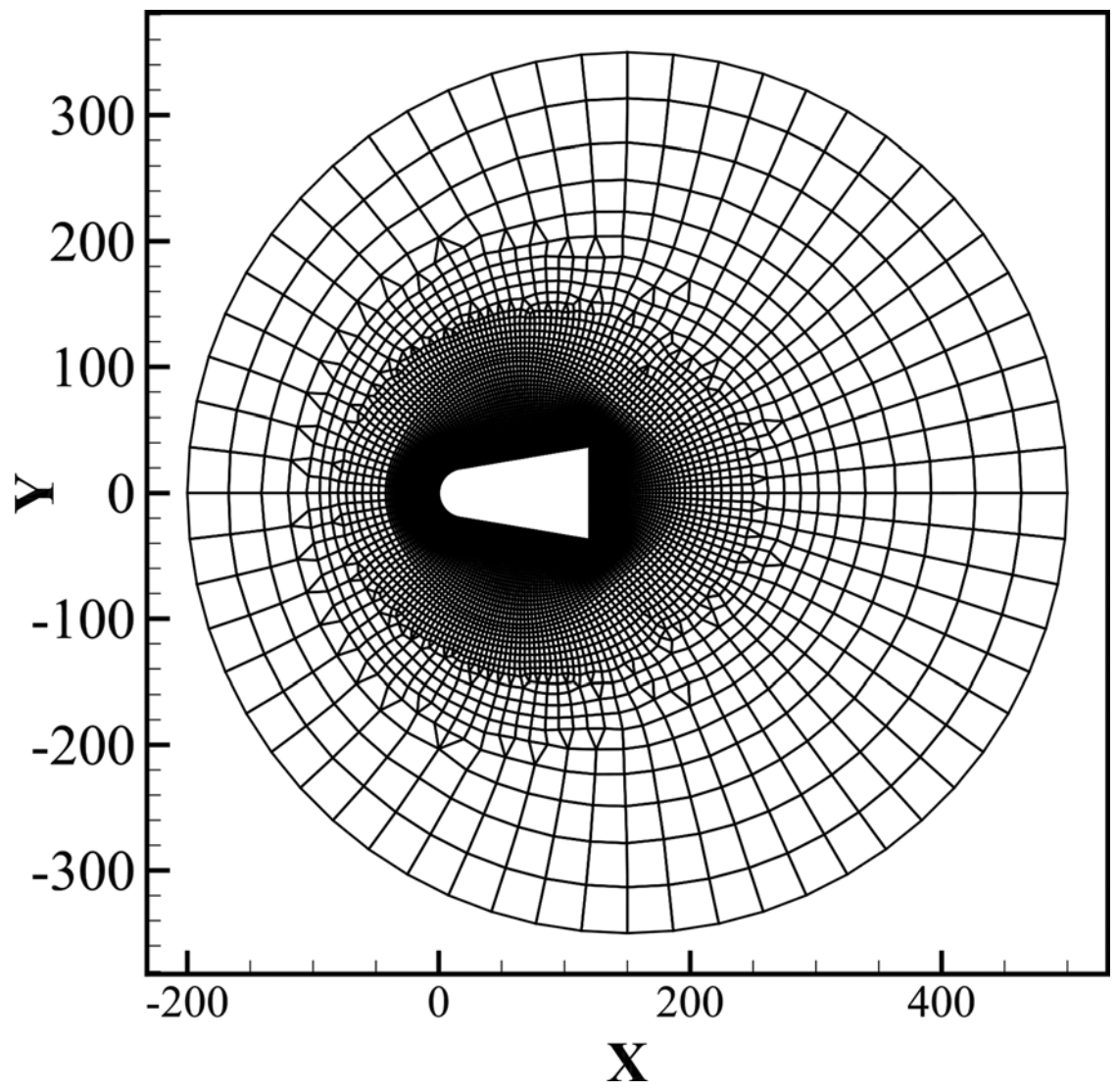

FIG. 10. Physical mesh for the blunt wedge simulation at Ma = 8.1 and Kn = 0.338.

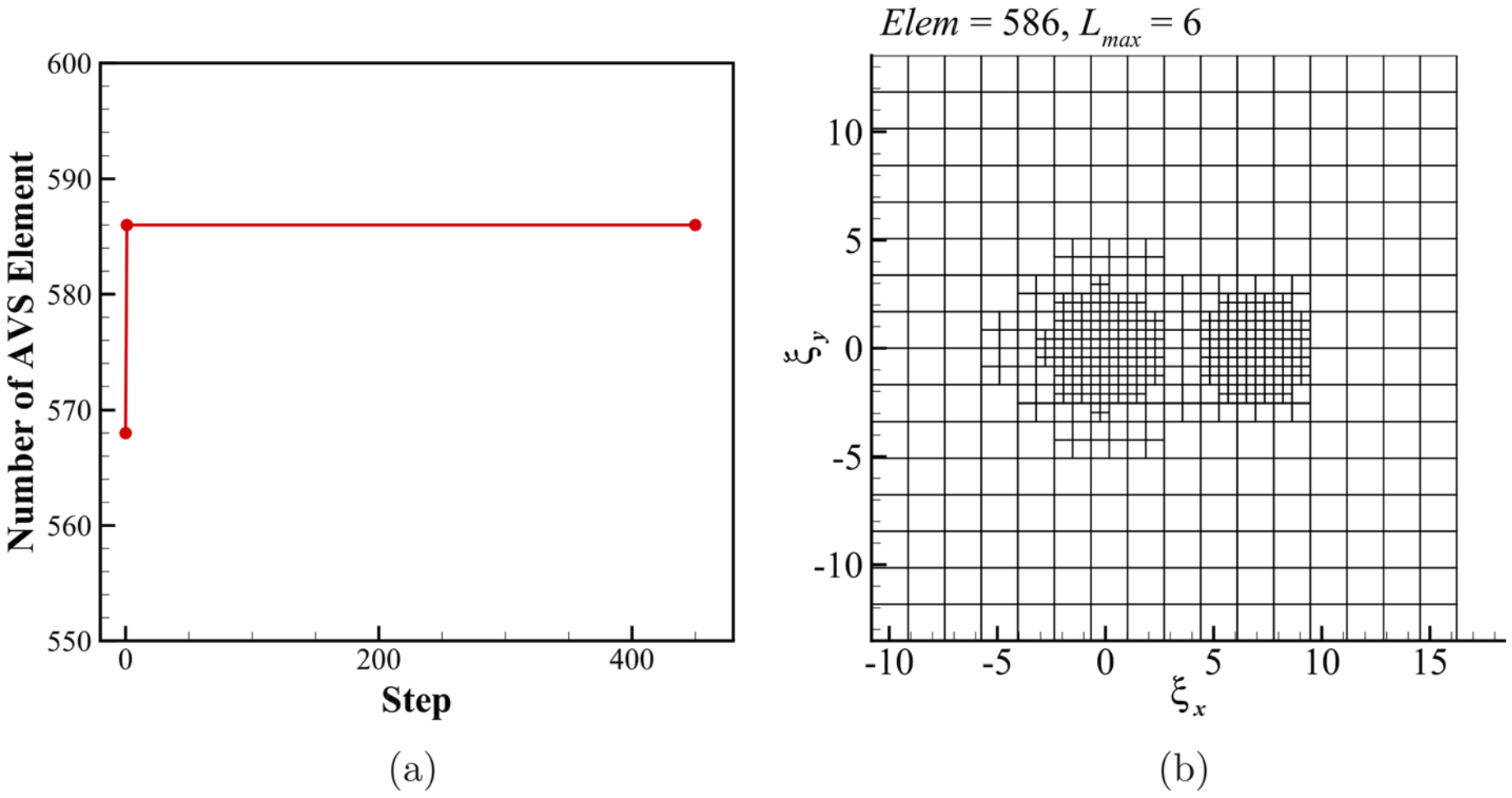

**FIG. 11.** AVS for the blunt wedge simulation at Ma = 8.1 and Kn = 0.338. (a) Evolution of the number of AVS elements, and (b) the final AVS.

the merging and splitting steps are illustrated using quadtree data for the 2 D case, with the 3 D case following a similar process. Figure 2(a) provides a concise diagram of a quadtree structure. The root node at level 0 serves as the root of the quadtree. When it splits, it generates four children ($k_1 - k_4$) at level 1. If a child node like $k_4$ needs to split again, it produces four more children ($k_{41} - k_{44}$) at level 2. The root node is the parent for nodes $k_1 - k_4$, while node $k_4$ is the parent for nodes $k_{41} - k_{44}$. Nodes that share the same parent are sibling nodes, and nodes with no children are termed leaf nodes. In Fig. 2(a), the leaf nodes are connected by a red line, and the corresponding velocity mesh is shown in Fig. 2(b). Notably, the velocity mesh includes only the leaf nodes $k_1$, $k_2$, $k_3$, $k_{41}$, $k_{42}$, $k_{43}$, and $k_{44}$. The merging process, essentially the reverse of the splitting process, occurs when all four sibling nodes are leaf nodes, and the merging conditions are met. For instance, the leaf nodes $k_{41} - k_{44}$ can only merge into the node $k_4$ when all of them satisfy the merging conditions. Then, the parent node $k_4$ becomes a leaf node.

After the adaptation procedure, it is necessary to map the distribution function from the old velocity mesh to the new velocity mesh. In this study, the reconstruction of the distribution function is accomplished using a unique direct mapping method instead of interpolation. Specifically, the newly generated leaf nodes share the same distribution function as their root, while the distribution function value for root nodes, whose leaves will be merged, is the average of the distribution function of the four leaves. Without loss of generality, Fig. 2 is employed to illustrate the reconstruction of the distribution function. For example, when the root node splits into four child nodes $k_1 - k_4$, the distribution functions of nodes $k_1 - k_4$ are identical to the distribution function of the root node,

$$f'_{k_i} = f_{root}, \quad i = 1, 2, 3, 4, \tag{17}$$

where $f$ and $f'$ are the distribution functions of the old velocity mesh and the new velocity mesh, respectively. In the merging step, the distribution function of the parent node is calculated as the average of the distribution function of its child nodes. For instance, when the four child nodes $k_{41} - k_{44}$ merge into their parent node $k_4$, the distribution function of node $k_4$ is obtained by averaging the distribution functions of nodes $k_{41} - k_{44}$,

$$f'_{k_4} = \frac{1}{4}\sum_{i=1}^{4} f_{4i}. \tag{18}$$

Notice that for conventional information mapping on adaptive Cartesian mesh, the distribution function for newly generated children is reconstructed using an interpolation method. However, for AVS, the distribution function is often non-smooth and may even be discontinuous in the velocity space. In such cases, interpolation may compromise the robustness of the numerical method. Therefore, in this work, the low-order direct mapping method described above is adopted. However, both direct mapping and interpolation mapping can affect the conservation properties due to their truncation errors. Therefore, forced conservation should be adopted as described in Sec. II B 2.

### 2. Conservation correction

In the reconstruction of distribution function, it is crucial to satisfy the conservation constraint. The mass conservation is upheld for the distribution function $f'$ obtained from the direct mapping method, while the momentum and energy conservation are not preserved. Following the approach outlined in the literature,[28,38] the forced conservation is employed to address this issue. By taking the moments of $f'$, we can obtain the corresponding macroscopic variable $\boldsymbol{W}'$ as

$$\boldsymbol{W}' = \int \psi f' d\xi. \tag{19}$$

The forced conservation is realized by using the equilibrium distribution, whose mathematical expression is simple and has no ambiguity for this fix at the continuum limit. The fixed distribution function on the new velocity mesh can be expressed as

$$f_{fix} = f' + g^{eq}(\boldsymbol{W}) - g^{eq}(\boldsymbol{W}'), \tag{20}$$

**TABLE IV.** Computational efficiency of the blunt wedge simulation.

| ... | CVS | AVS | Ratio |
|---|---|---|---|
| Elements | 7 921 | 586 | 13.5 |
| Steps | 430 | 450 | 0.96 |
| Time (s) | 15 622.9 | 1499.3 | 10.4 |

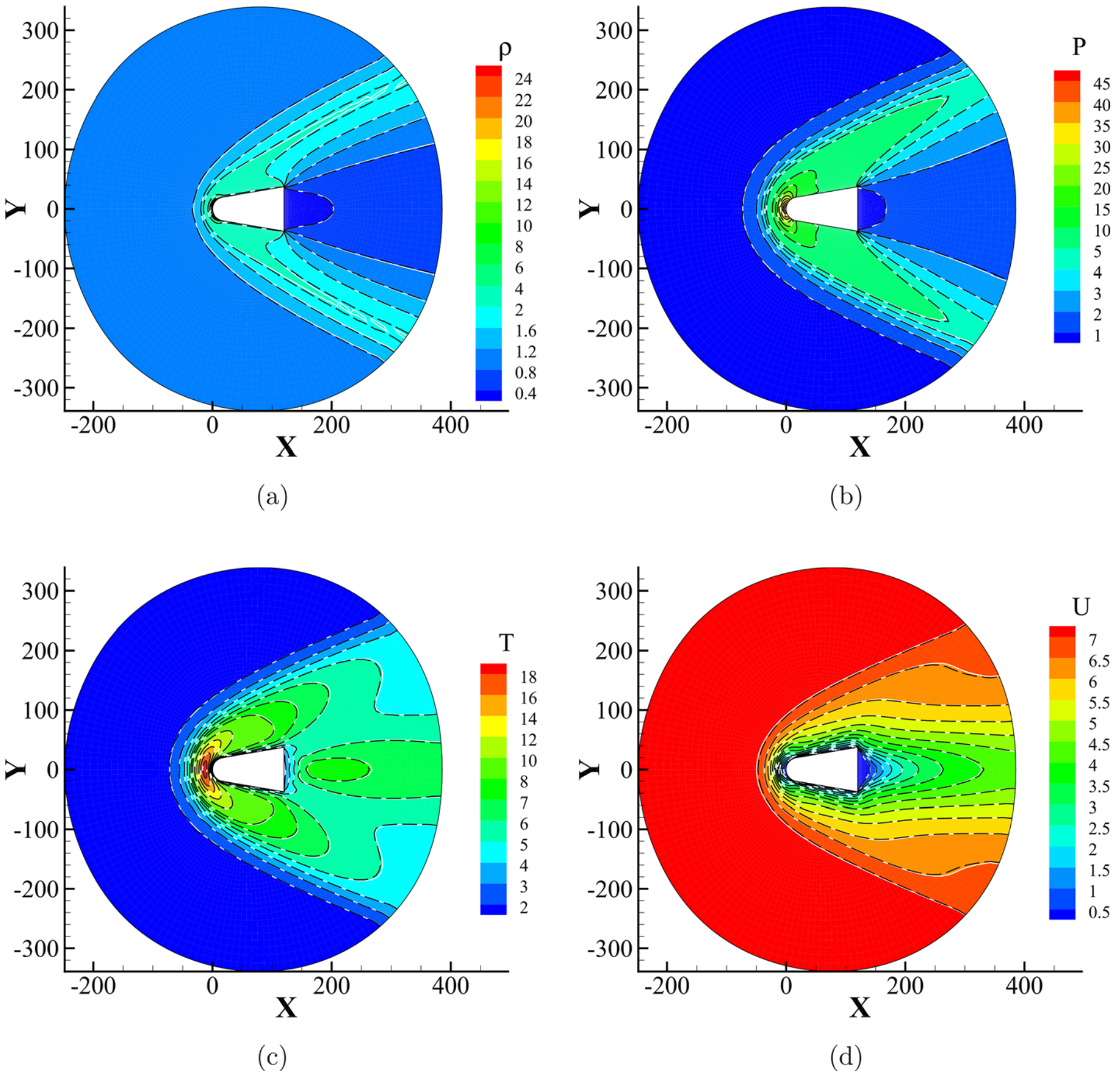

**FIG. 12.** Contours around the blunt wedge (background and white solid lines: CVS; black long dashed line: AVS). (a) Density, (b) pressure, (c) temperature, and (d) horizontal velocity.

where $g^{eq}(\boldsymbol{W})$ and $g^{eq}(\boldsymbol{W}')$ are the Maxwellian distribution functions corresponding to macroscopic variables $\boldsymbol{W}$ and $\boldsymbol{W}'$, respectively. Taking moments of the distribution function $f_{fix}$ in Eq. (20), one can obtain

$$\int \boldsymbol{\psi} f_{fix} d\boldsymbol{\xi} = \boldsymbol{W}' + \boldsymbol{W} - \boldsymbol{W}' = \boldsymbol{W}. \tag{21}$$

Therefore, the macroscopic variables of the new velocity mesh remain consistent with those of the old velocity mesh, ensuring the preservation of mass, momentum, and energy conservation. Furthermore, for weakly nonequilibrium flows, the distribution function of a new velocity mesh can be directly set as the equilibrium distribution function corresponding to the macroscopic variables of the old velocity mesh

$$f = g^{eq}(\boldsymbol{W}). \tag{22}$$

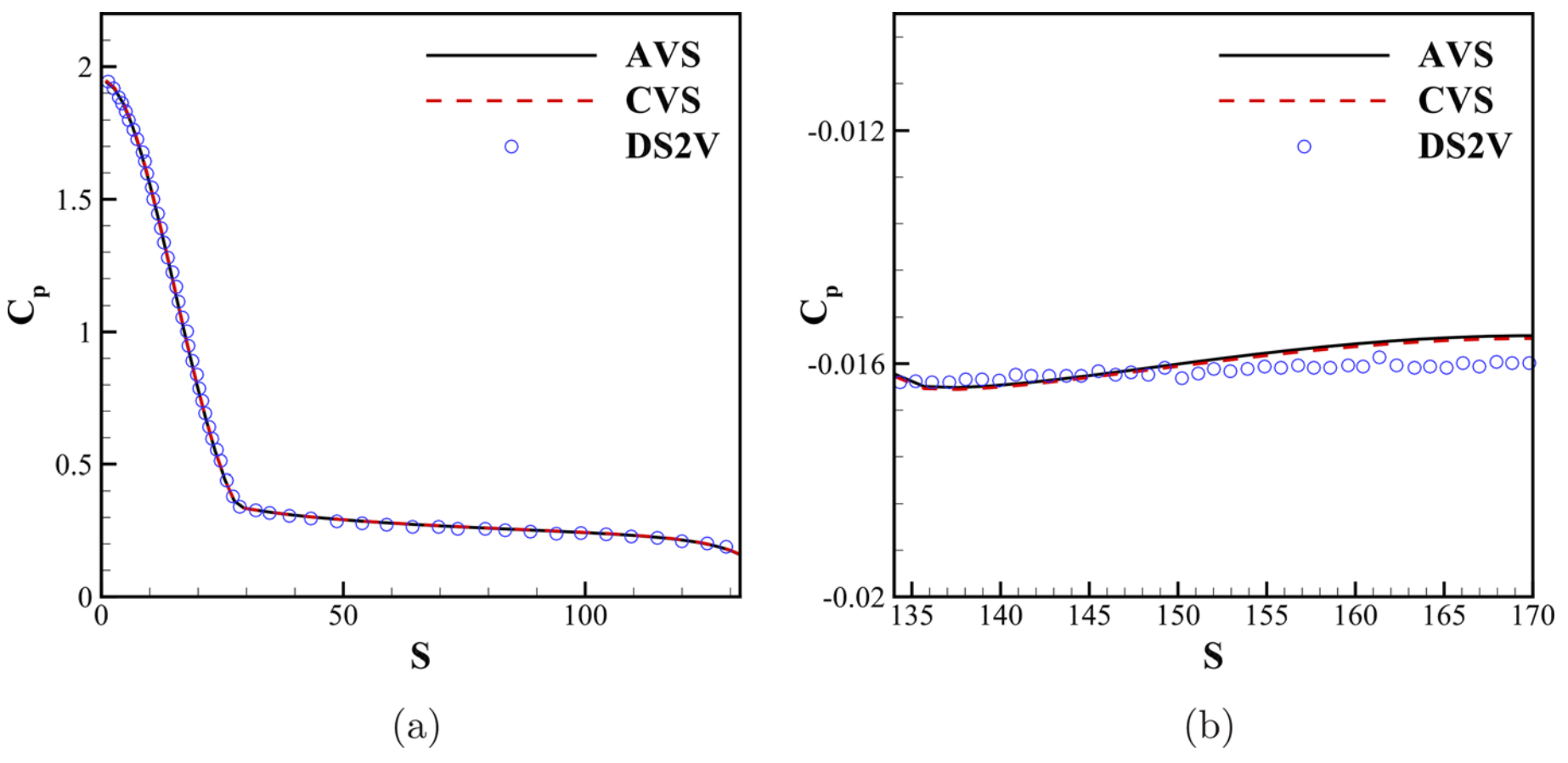

**FIG. 13.** Pressure coefficient on the surface of blunt wedge at Ma = 8.1 and Kn = 0.338. (a) Body and (b) bottom.

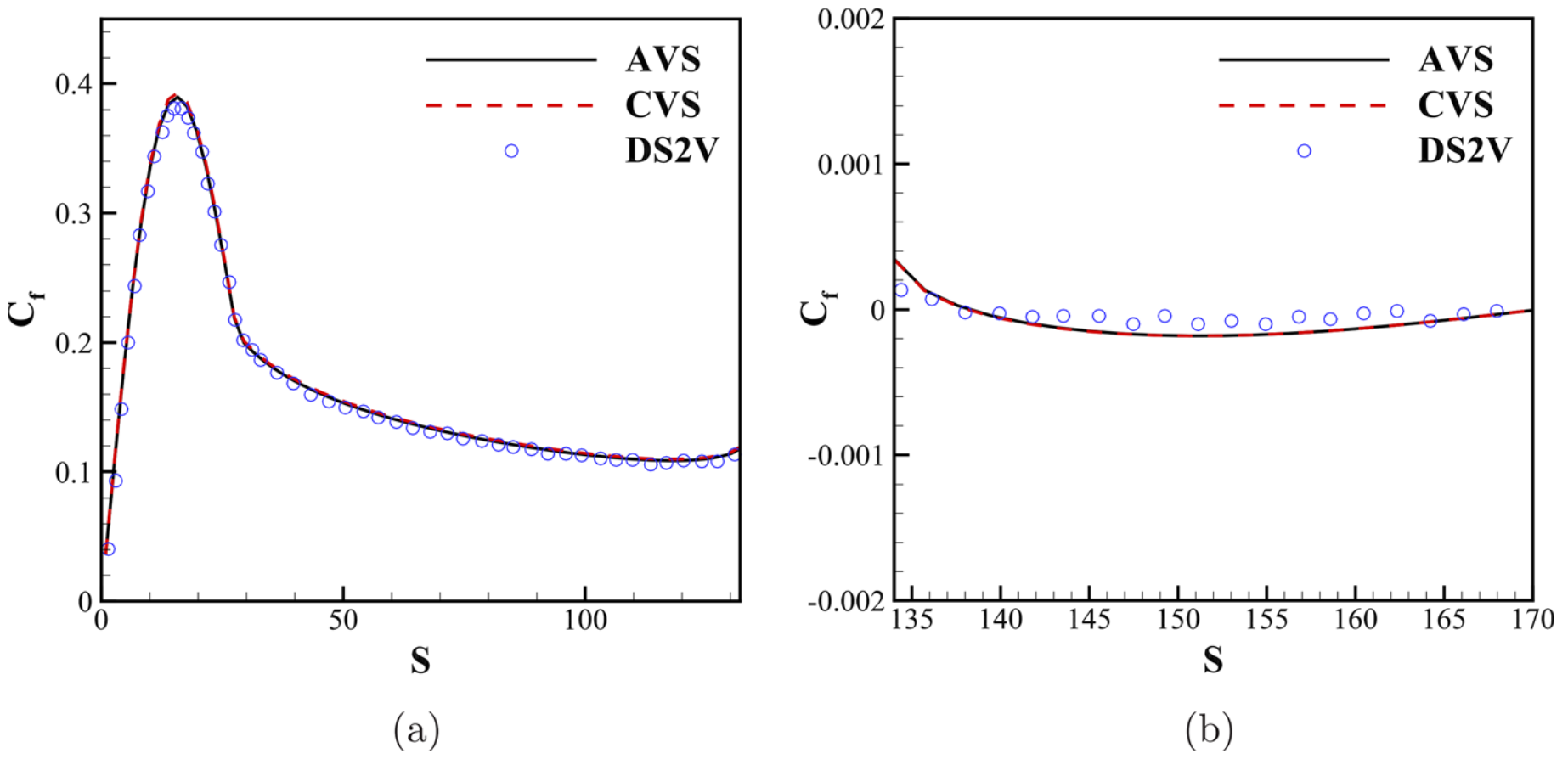

FIG. 14. Skin friction coefficient on the surface of blunt wedge at Ma = 8.1 and Kn = 0.338. (a) Body and (b) bottom.

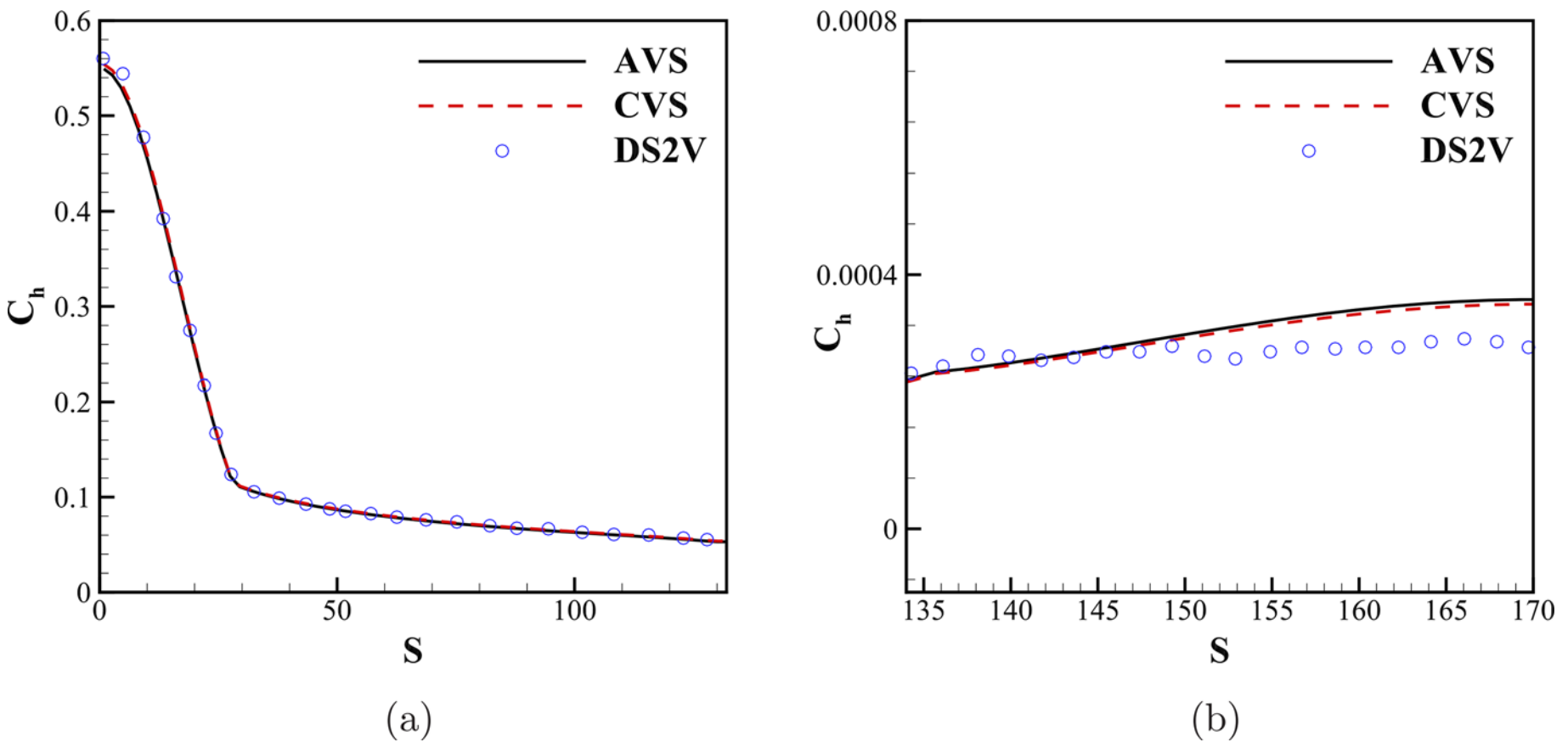

FIG. 15. Heat transfer coefficient on the surface of blunt wedge at Ma = 8.1 and Kn = 0.338. (a) Body and (b) bottom.

### 3. The domain of AVS and the multi-zone strategy to improve the performance

In this section, the optimized strategy of global AVS is proposed and discussed. The performance of AVS in capturing flow behavior hinges on its domain and resolution. Therefore, the AVS should cover all the effective molecular velocities. For example, in hypersonic flow passing a cylinder with a Mach ($Ma$) number of 10, molecules often concentrate within two small regions centered at the macroscopic speed of the freestream and the zero-velocity near the cylinder, respectively. Therefore, the center of the domain of global AVS is not at the zero-velocity point. Furthermore, some sub-zones are automatically created at these concentrated regions of molecules, and their adaptation levels are suggested for better performance. For some cases, after the sub-zones are created and the suggested adaptations are made, the AVS may not even need any further adaptation. Figure 3 illustrates the diagram of the global AVS. The radius of the velocity domain is determined by $R_{dv} = a\sqrt{RT_0}$, where $T_0$ denotes the total temperature and $a$ ranges between 4 and 7. The center of the velocity domain is positioned at $0.4\boldsymbol{U}$ to cover three refined zones (Table I): the freestream zone (zone A), the stationary zone (zone B), and the separated zone (zone C). The freestream zone has a radius of $R_A = 3\sqrt{RT_\infty}$ and a center at $\boldsymbol{U}_\infty$. The stationary zone, however, has a radius of $R_B = 3\sqrt{RT_w}$ and its center at zero. Finally, the separated zone, which is intended for separate flow, has a radius of $R_C = 5\sqrt{RT_w}$ and a center at zero. Here, $T_\infty$ is the freestream temperature and $T_w$ is the wall temperature.

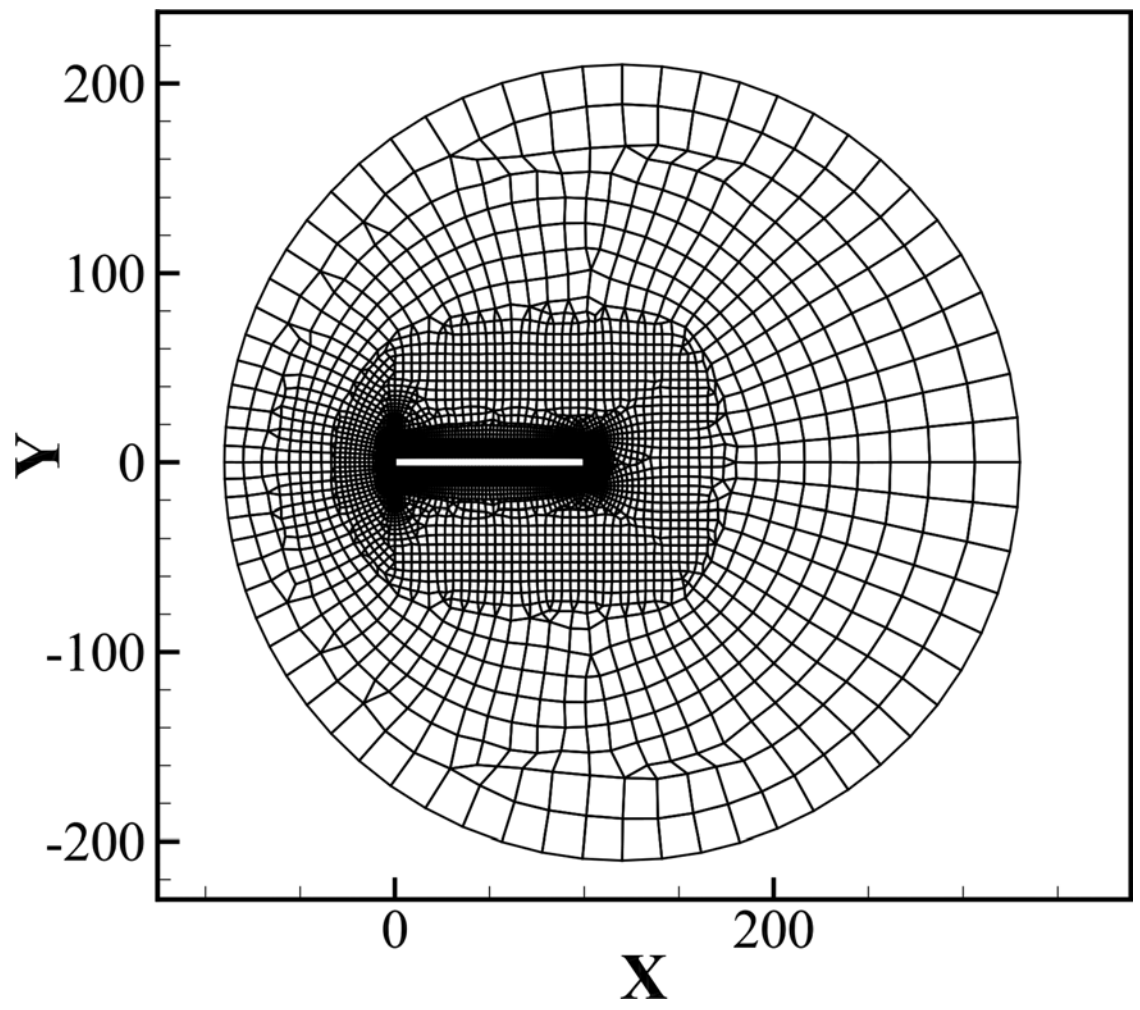

FIG. 16. Physical mesh for the truncated flat plate simulation at Ma = 20.2 and Kn = 0.0169.

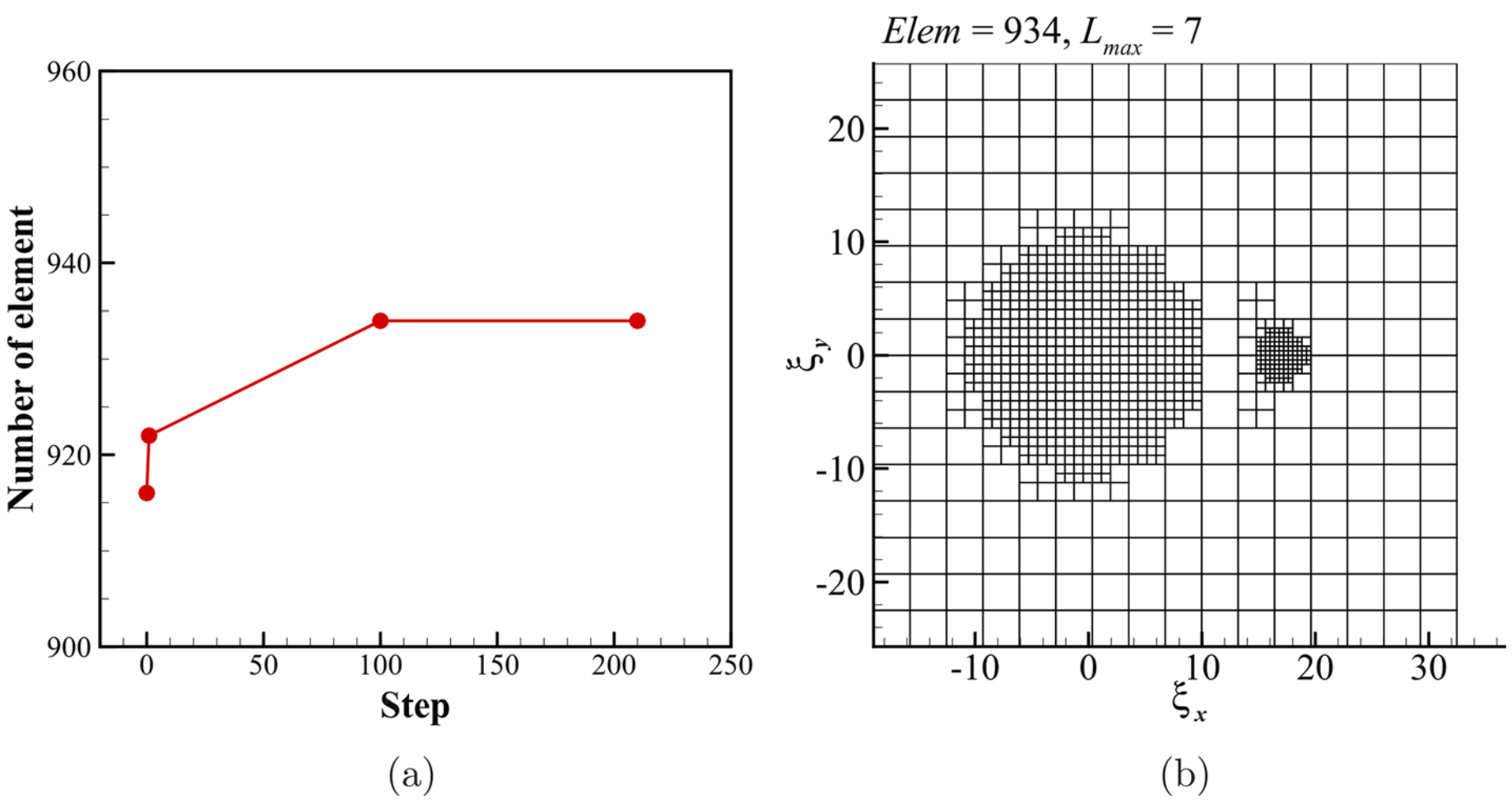

FIG. 17. AVS for the truncated flat plate simulation with AOA = 0°. (a) Evolution of the number of AVS elements and (b) the final AVS.

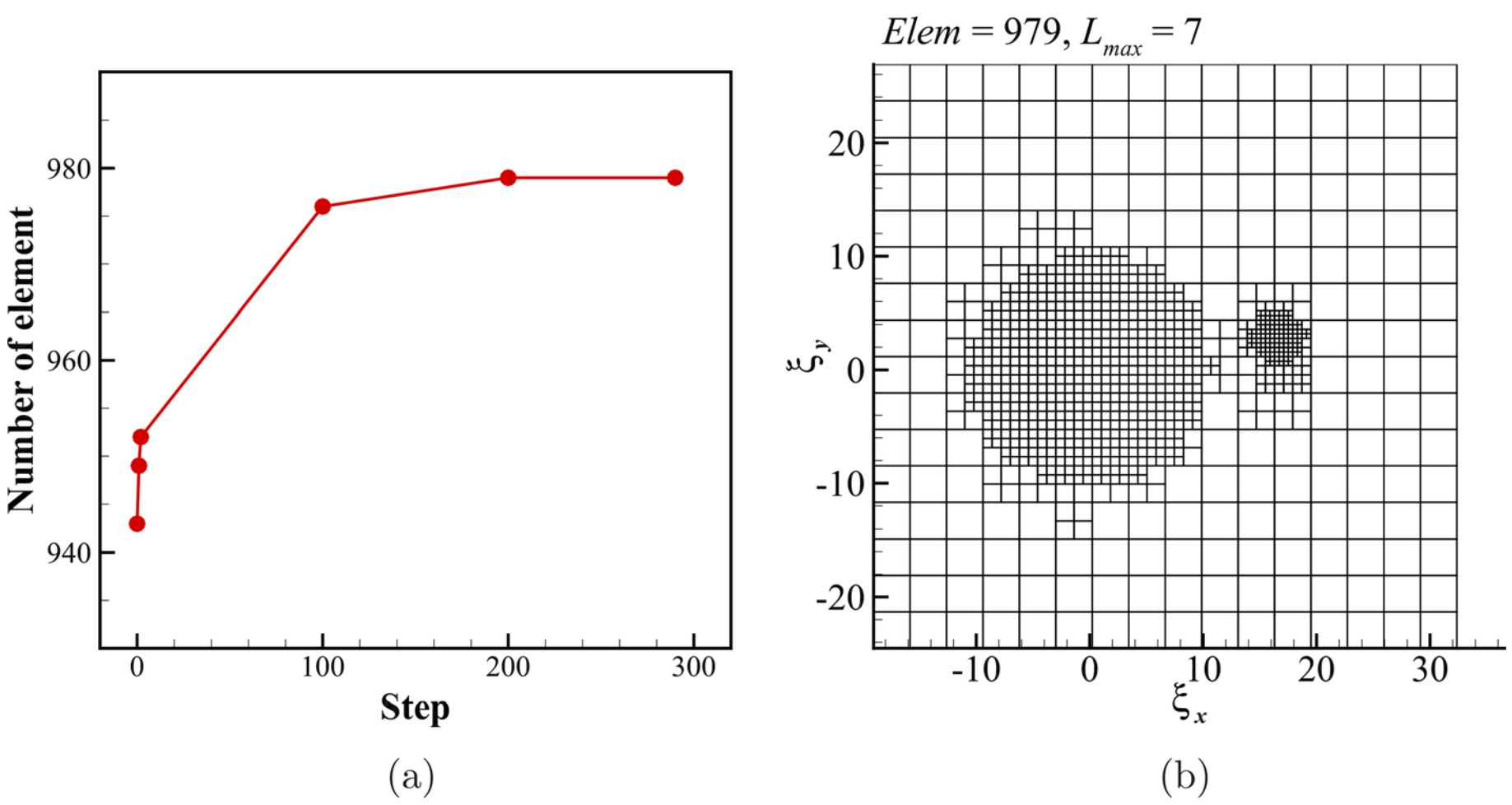

FIG. 18. AVS for the truncated flat plate simulation with AOA = 10°. (a) Evolution of the number of AVS elements and (b) the final AVS.

To ensure the resolution of these three refined zones, their respective levels are determined. Typically, the freestream zone requires the highest resolution and the maximum level,

$$L_A = L_{max}, \tag{23}$$

where $L_A$ represents the level of the freestream zone, the maximum level $L_{max}$ will be discussed later. Once the maximum level $L_{max}$ has been determined, the level of the stationary zone can be estimated as

$$L_B = \begin{cases} L_{max}, & \dfrac{T_w}{T_\infty} < 4; \\ L_{max} - 1, & otherwise. \end{cases} \tag{24}$$

TABLE V. Computational efficiency of the truncated flat plate simulation (AOA = 0°).

| ... | CVS | AVS | Ratio |
|---|---|---|---|
| Elements | 10 201 | 934 | 10.9 |
| Steps | 210 | 210 | 1.0 |
| Time (s) | 9 542.1 | 971.0 | 9.8 |

TABLE VI. Computational efficiency of the truncated flat plate simulation (AOA = 10°).

| ... | CVS | AVS | Ratio |
|---|---|---|---|
| Elements | 10 201 | 979 | 10.4 |
| Steps | 330 | 290 | 1.1 |
| Time (s) | 14 765.7 | 1395.6 | 10.6 |

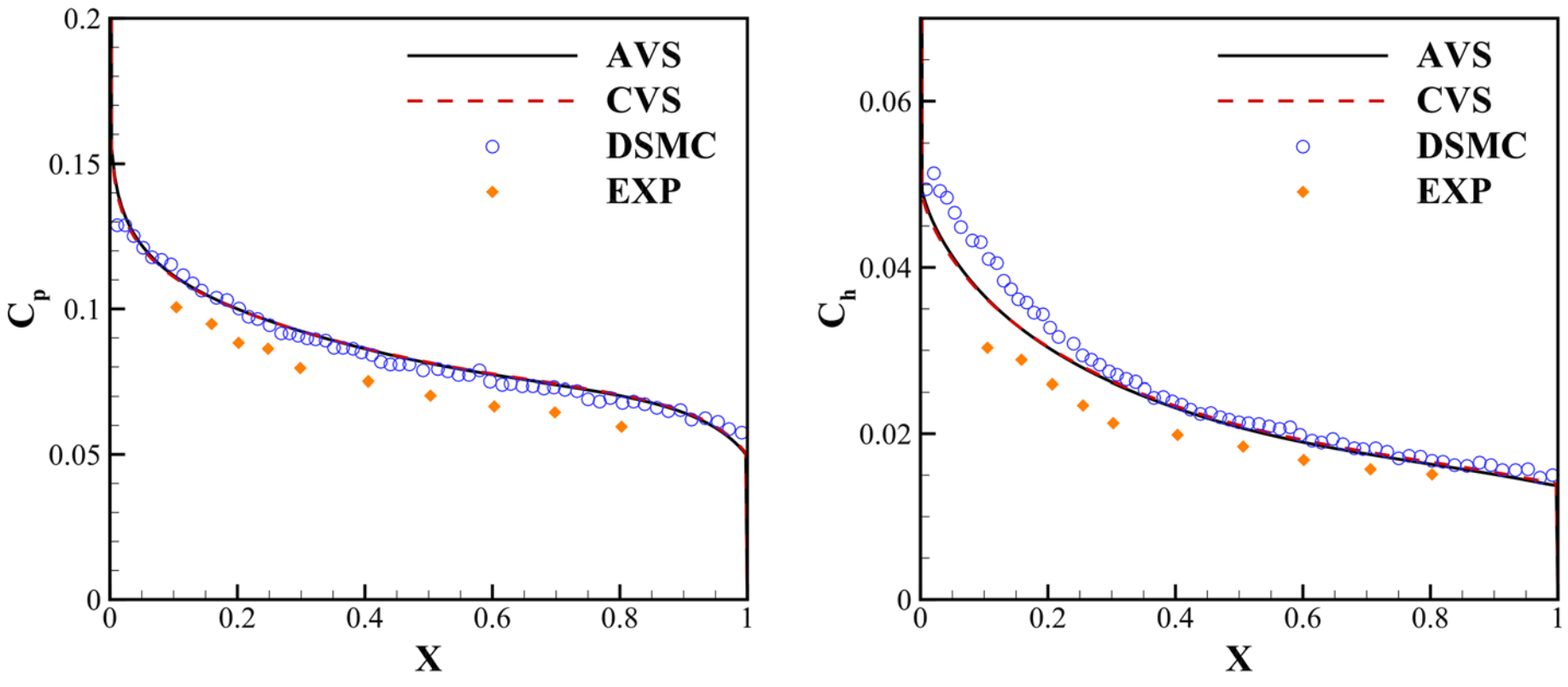

(a) (b)

FIG. 19. Physical variables on the lower surface of truncated flat plate at AOA = 0° (Ma = 20.2, Kn = 0.0169). (a) Pressure coefficient and (b) heat transfer coefficient.

The level of AVS outside the refined zone falls within the range of the minimum level $L_{min}$ and the maximum level $L_{max}$, with the minimum level being

$$L_{min} = \begin{cases} 4, & Ma > 5; \\ 3, & otherwise. \end{cases} \tag{25}$$

Additionally, if there is a separated flow, the separated zone should be refined, and its level is determined by

$$L_C = max(L_{min}, L_B - 1), \quad 3\sqrt{RT_w} < |\xi| < 5\sqrt{RT_w}, \tag{26}$$

where $|\xi|$ is the magnitude of the discrete velocity.

Now, let us focus on the maximum level $L_{max}$ of the AVS. In general, it is feasible to predict the approximate radius $R_{dv}$ and minimum element size $h_{min}$ of the velocity mesh, but predicting the maximum level is more challenging. Additionally, since the maximum level is an integer, adjusting either the radius or the minimum element size will affect the other. For instance, if the radius is $R_{dv}$ and the approximate minimum element size is $h_{min,apr}$, the corresponding $L_{max,apr}$ can be determined as

$$2^{L_{max,apr}-1} \geq \frac{R_{dv}}{h_{min,apr}}. \tag{27}$$

The coefficient 2 on the left side of Eq. (27) represents the ratio between the element size of two adjacent levels. Therefore, the maximum level is

$$L_{max} = min(L_{max,apr}). \tag{28}$$

The final minimum element size is

$$h_{min} = \frac{R_{dv}}{2^{L_{max}-1}}. \tag{29}$$

It is evident that $h_{min}$ does not exceed $h_{min,apr}$ and is also not less than half of $h_{min,apr}$. On the contrary, if the minimum element size is set to $h_{min}$ and the approximate radius is $R_{dv,apr}$, the corresponding $L_{max,apr}$ can be determined as

$$2^{L_{max,apr}-1} \geq \frac{R_{dv,apr}}{h_{min}}. \tag{30}$$

Similarly, the maximum level is

$$L_{max} = min(L_{max,apr}). \tag{31}$$

The final radius is

$$R_{dv} = (2^{L_{max}-1})h_{min}. \tag{32}$$

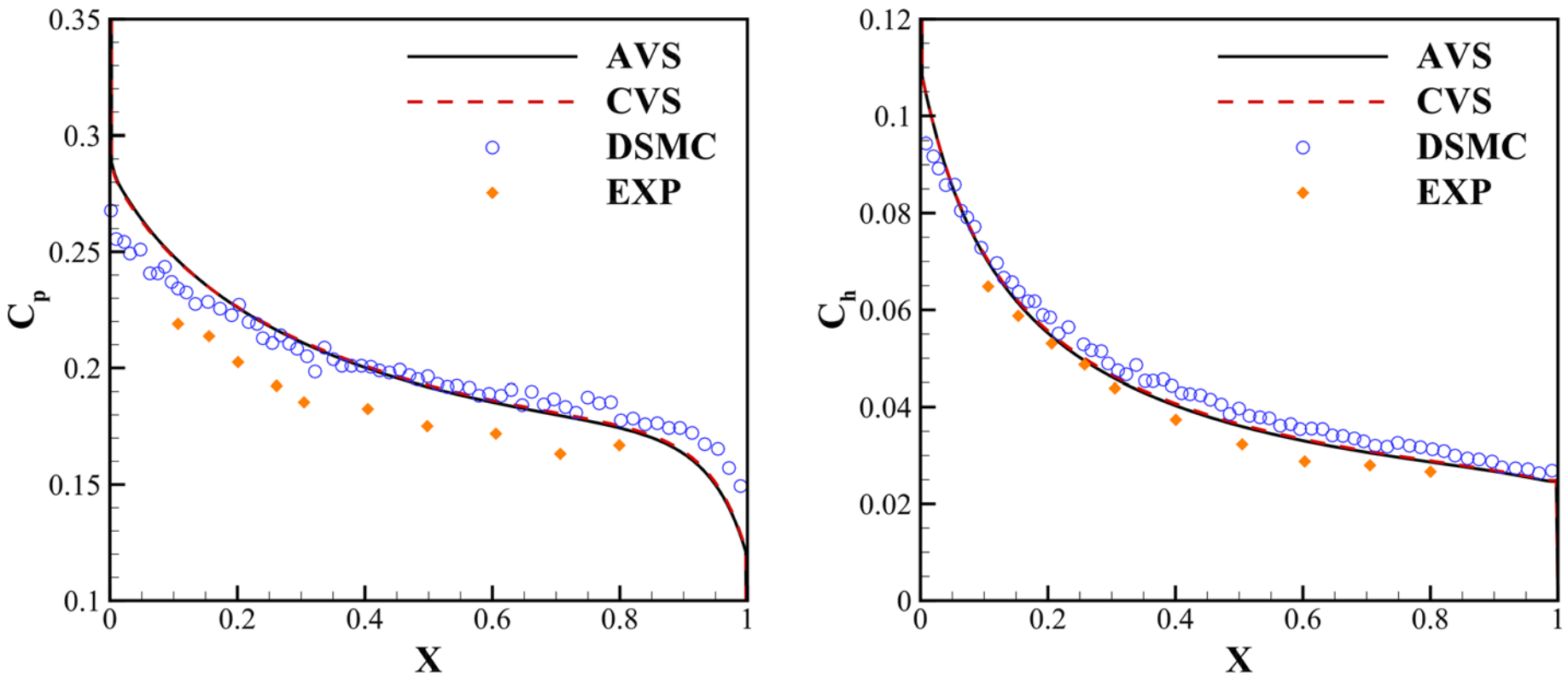

(a) (b)

FIG. 20. Physical variables on the lower surface of truncated flat plate at AOA = 10° (Ma = 20.2, Kn = 0.0169). (a) Pressure coefficient and (b) heat transfer coefficient.

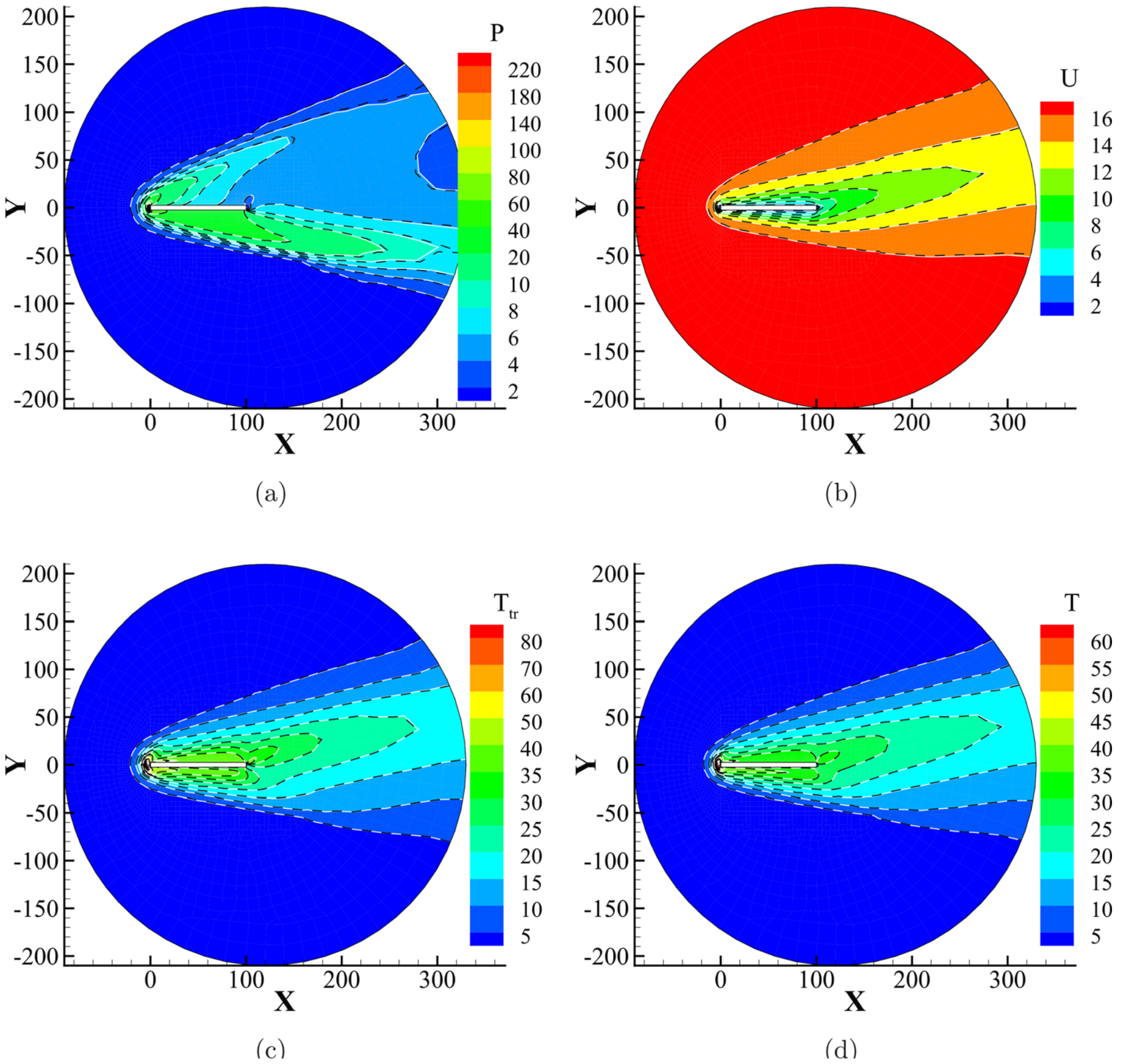

FIG. 21. Contours around the truncated flat plate at AOA = 10° (background and white solid lines: CVS; black long dashed line: AVS). (a) Pressure, (b) horizontal velocity, (c) translational temperature, and (d) equilibrium temperature.

In summary, two methods are provided for determining $L_{max}$: Method 1 involves providing $R_{dv}$ with an approximate $h_{min,apr}$, and method 2 involves providing $h_{min}$ with an approximate $R_{dv,apr}$. In the work of Baranger *et al.*,[39,40] the minimum element size of the velocity mesh is determined as $\min(\alpha\sqrt{RT_j})$, where $j$ represents the index of the physical cell and $\alpha \leq 1.0$. In this paper, for method 1, the radius is fixed at $R_{dv} = 4\sqrt{RT_0}$, and $h_{min,apr}$ is set to 0.6 (equivalent to $0.8485\sqrt{RT_\infty}$). Method 1 is utilized for simulating vehicles in freestream conditions (external flows), such as flow over a cylinder, blunt wedge, and truncated flat plate. However, method 2 is employed for specialized simulations, including the 2 D Riemann problem, jet flow expanding into a vacuum environment, and flow over a sphere with high wall temperature.

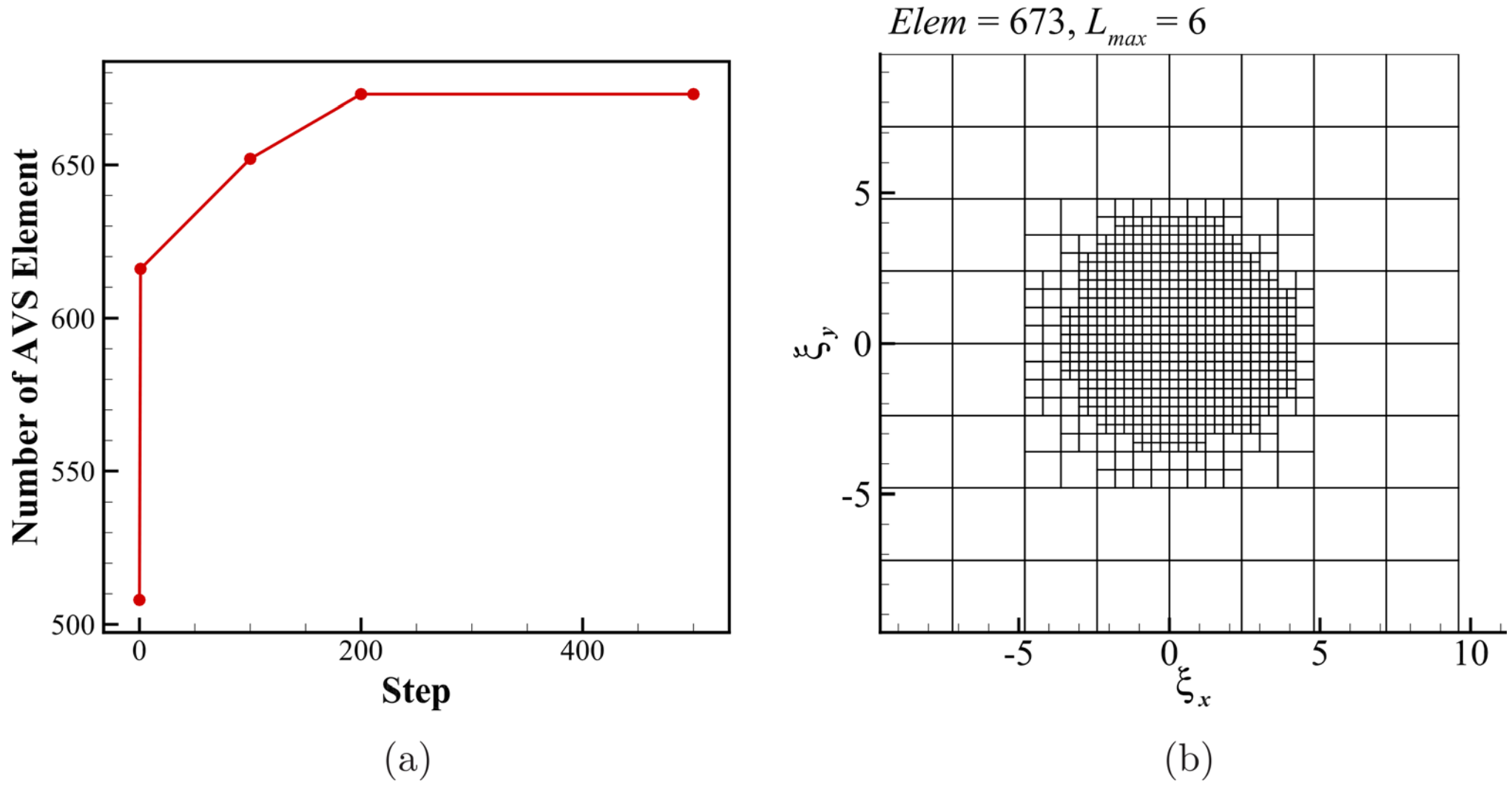

FIG. 22. AVS for the 2 D Riemann problem ($\mu_0 = 10$). (a) Evolution of the number of AVS elements and (b) the final AVS.

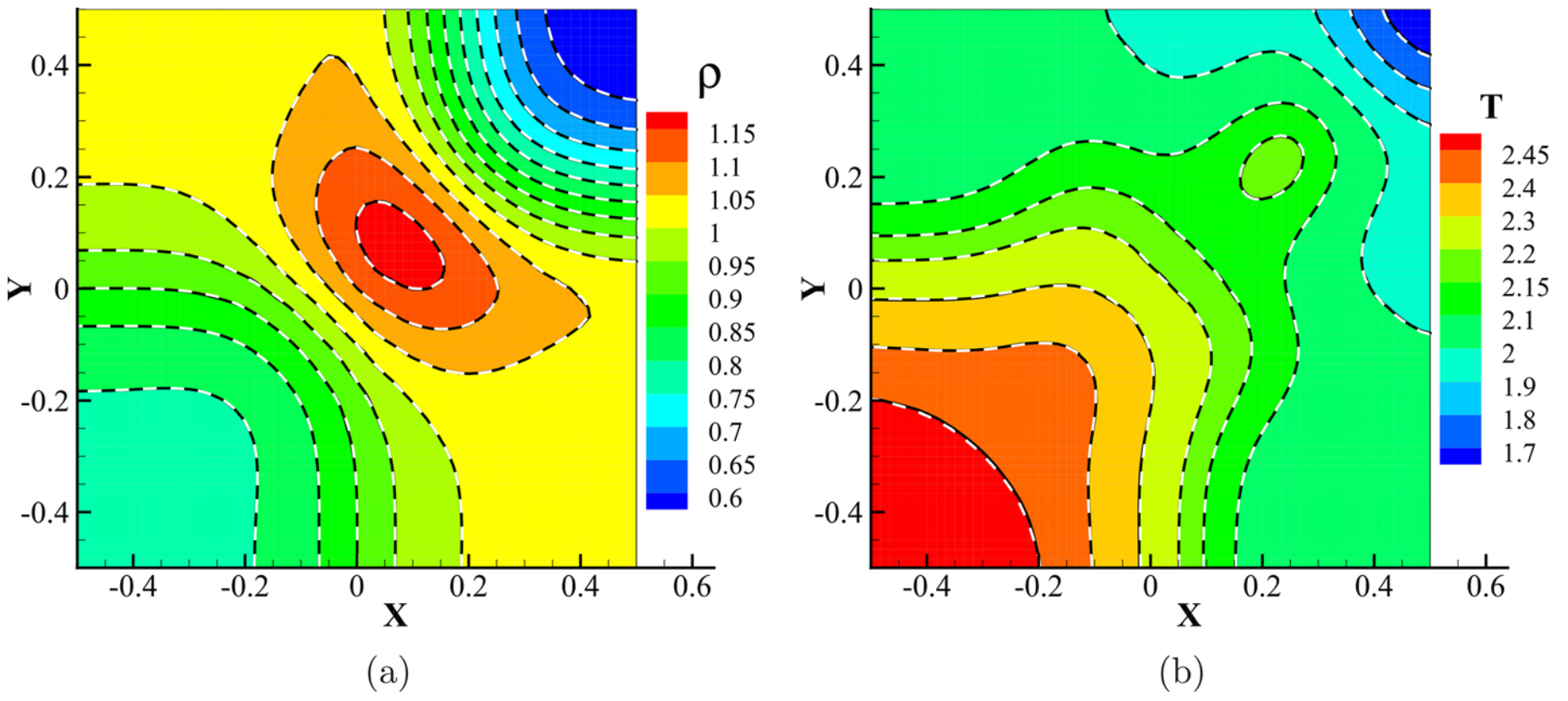

FIG. 23. Contours of the 2 D Riemann problem (background and white solid lines: CVS; black long dashed line: AVS). (a) Density and (b) temperature.

Once all the adaptive parameters are determined, an initial AVS can be created

(1) Create an element with only a root node, whose level is 0.
(2) Split the single element $N$ times to generate a unified mesh of $2^N \times 2^N$ elements, where $N$ can be set to $L_w - 1$.
(3) Perform one adaption based on the physical variables and equilibrium distribution function of the freestream zone and stationary zone.

As a result, an initial AVS is obtained. The flow chart of the present method is depicted in Fig. 4. To ensure effective alignment with the initial flow field, adaptation is carried out during the first three steps. Subsequently, adaptation occurs at regular intervals, usually every 10 or 50 steps.

## III. NUMERICAL EXPERIMENTS

In this section, several test cases are conducted to validate the proposed method. The simulation results from the DS2V software[1] and literature data are used as reference benchmarks, and the results of AVS and CVS are compared to these benchmarks. Since our solver is a parallel program, we require more than one process for computation. Therefore, two processes are utilized for 2 D simulations to compare the computational efficiency of AVS and CVS.

Generally, four independent characteristic variables are introduced in the non-dimensional reference system, namely, the reference length $L_{ref} = L_c$, the reference temperature $T_{ref} = T_\infty$, the reference density $\rho_{ref} = \rho_\infty$, and the reference speed $U_{ref} = \sqrt{2RT_{ref}}$, where $L_c$ is the characteristic length scale of the flow, and $T_\infty$ and $\rho_\infty$ are the temperature and density of the freestream, respectively. Thus, the following basic non-dimensional quantities can be obtained:

$$\hat{L} = \frac{L}{L_{ref}}, \quad \hat{T} = \frac{T}{T_{ref}}, \quad \hat{\rho} = \frac{\rho}{\rho_{ref}}, \quad \hat{U} = \frac{U}{U_{ref}}. \tag{33}$$

One can obtain a complete non-dimensional system by employing these basic quantities. Unless declared otherwise, all variables in the following that lack a "hat" are non-dimensional quantities for simplicity's sake.

TABLE VII. Computational efficiency of the 2 D Riemann problem.

| ... | CVS | AVS | Ratio |
|---|---|---|---|
| Elements | 10 201 | 673 | 15.2 |
| Steps | 500 | 500 | 1.0 |
| Time (s) | 3 257.6 | 247.8 | 13.1 |

### A. Hypersonic flow over a cylinder

The relationship between $C_1$ and $Ma$ is explored through the simulations of hypersonic flow over a cylinder with a wide range of Mach numbers. The Knudsen ($Kn$) number based on the radius of the cylinder is 1.0. The working gas is Argon and the variable hard-sphere (VHS) model with $\omega = 0.81$ is employed. The temperature of both the freestream and the surface of the cylinder is 273 K. In this study, the radius of the cylinder serves as the reference length, while the density and temperature of the freestream serve as the reference density and temperature. Consequently, both the dimensionless density and temperature are 1.0. The dimensionless radius of the cylinder is 1.0.

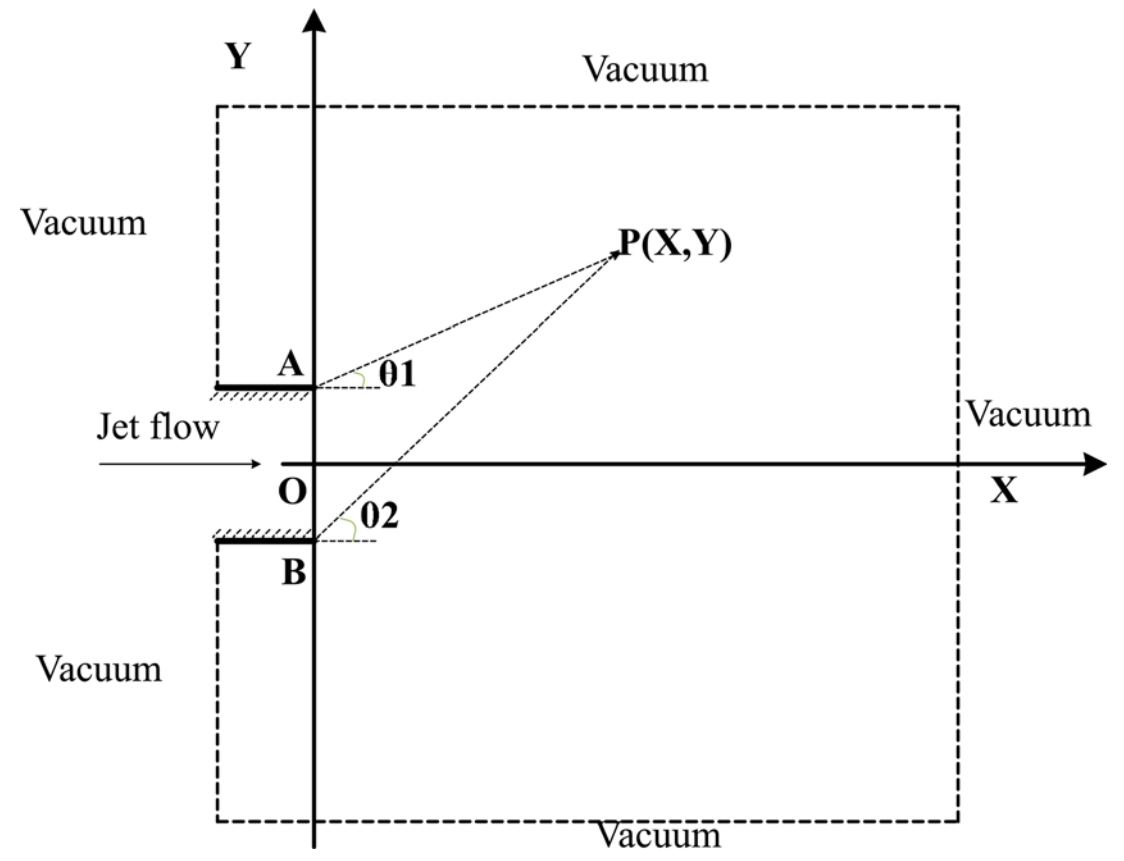

FIG. 24. Geometry of the jet flow.

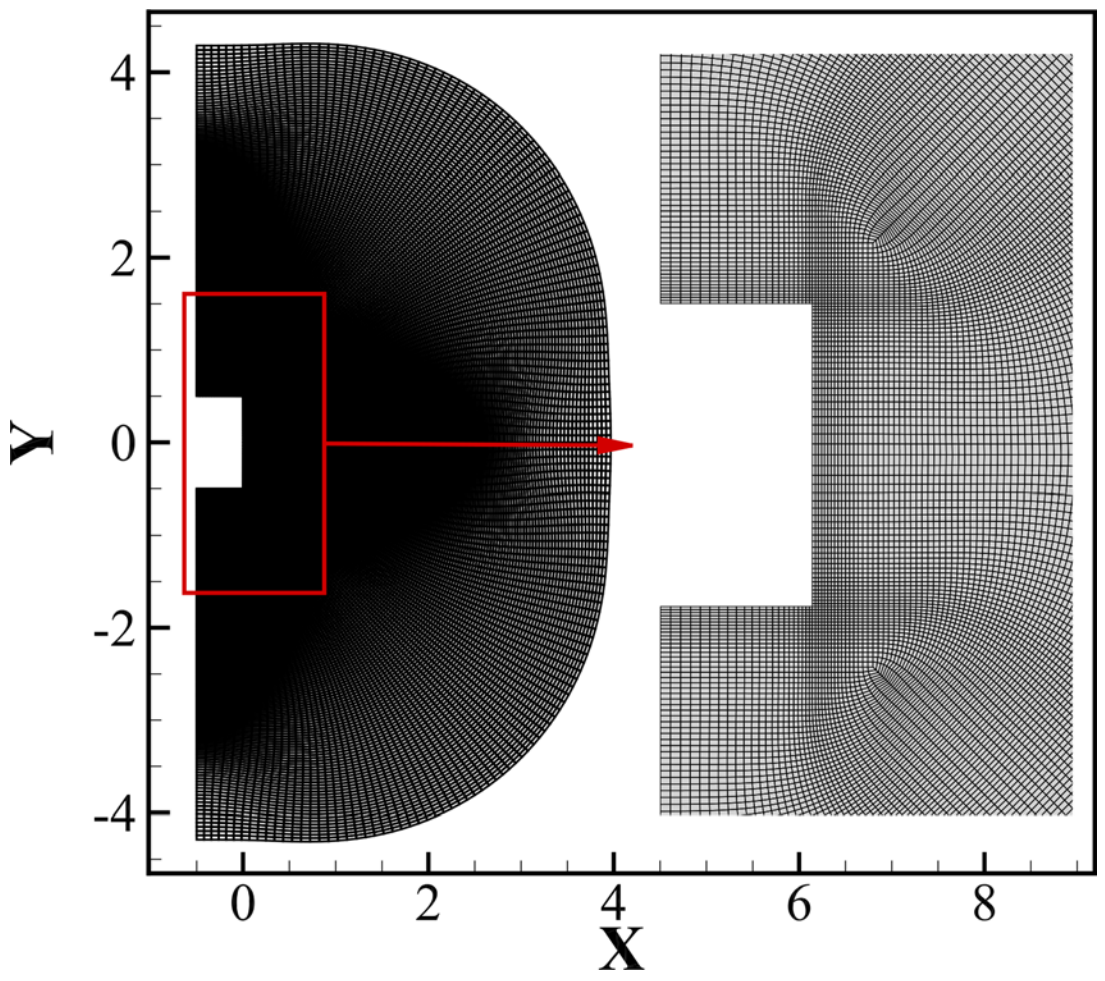

FIG. 25. Physical mesh of the jet flow.

The physical domain, which is a circular region centered at (0,0) with a radius of 15, is discretized by a mesh consisting of 64 × 61 cells. To accurately capture the heat flux on the cylinder surface, the height of the first layer of the mesh is set to 0.01. Both CVS and AVS are employed in this simulation to compare the results. For Mach numbers ranging from 2 to 6, 8 to 20, and 22 to 30, the CVS is discretized into 89 × 89, 101 × 101, and 201 × 201 points, respectively. In this case, method 1 for calculating $L_{\max}$ is adopted. The details of the final AVS are presented in Table II, and the discrete points ($Ma$, $C_1$) are plotted in Fig. 5. Additionally, a curve is fitted to express the relationship between $C_1$ and $Ma$ (Fig. 5) as

$$C_1 = 13.749Ma^{-1.465}. \tag{34}$$

In this paper, the adaptive parameter $C_1$ will be determined by Eq. (34) when method 1 for calculating $L_{max}$ is adopted.

The hypersonic flow passing a cylinder with a Mach number of 5 and a Knudsen number of 1.0 is one of the most classified multi-scale flows. Figure 6(a) illustrates the evolution of the number of AVS elements, and Fig. 6(b) presents the final AVS with a maximum level of 5, comprising 232 elements. Although adaptation continues throughout the simulation, the AVS at step 1 is adequate for this case. The computational efficiency is displayed in Table III. It is evident that the computation time with AVS is 22.4 times faster than that with CVS. On the one hand, since the elements of CVS is 34.1 times that of AVS in each process, the parallel communication time of AVS constitutes a larger proportion than that of CVS, leading to lower parallel efficiency for AVS. On the other hand, the variable time step of the present implicit UGKS is determined by the macroscopic residual.[37] Consequently, variations in the integral accuracy among different DVS would result in discrepancies in the macroscopic residual, leading to a differing convergence iteration step. In this instance, the convergence iteration step for CVS and AVS are 230 and 240, respectively. Consequently, the acceleration of AVS is less than 34.1. Figure 7 illustrates the density, horizontal velocity (U), and temperature along the stagnation line in front of the cylinder. Additionally, Fig. 8 depicts the pressure coefficient, skin friction coefficient, and heat transfer coefficient on the surface of the cylinder. The results obtained from AVS, CVS, and DSMC exhibit good agreement, indicating that AVS can accurately capture the behavior of the cylinder flow with less computational cost.

## B. Hypersonic flow over a blunt wedge

To further demonstrate the performance of the proposed method and validate the relationship between $C_1$ and Ma [Eq. (34)], we simulate the hypersonic flow over a blunt wedge, referencing a configuration from the literature.[41] The geometry of the blunt wedge is depicted in Fig. 9, featuring a length of $L = 120$ mm, a head radius of $R = 20$ mm, and a bottom height of $H = 74.72$ mm. Here, the working gas is argon, and the VHS model with $\omega = 0.81$ is employed. The Mach number and angle of attack (AOA) of the freestream are 8.1 and 0°, respectively. The temperature of freestream and surface are 189 K (equivalent to an altitude of 85 km) and 273 K. The Knudsen numbers based on $R$ and $H$ are 0.338 and 0.090, respectively. The unstructured physical mesh, consisting of 11 905 cells, is depicted in Fig. 10.

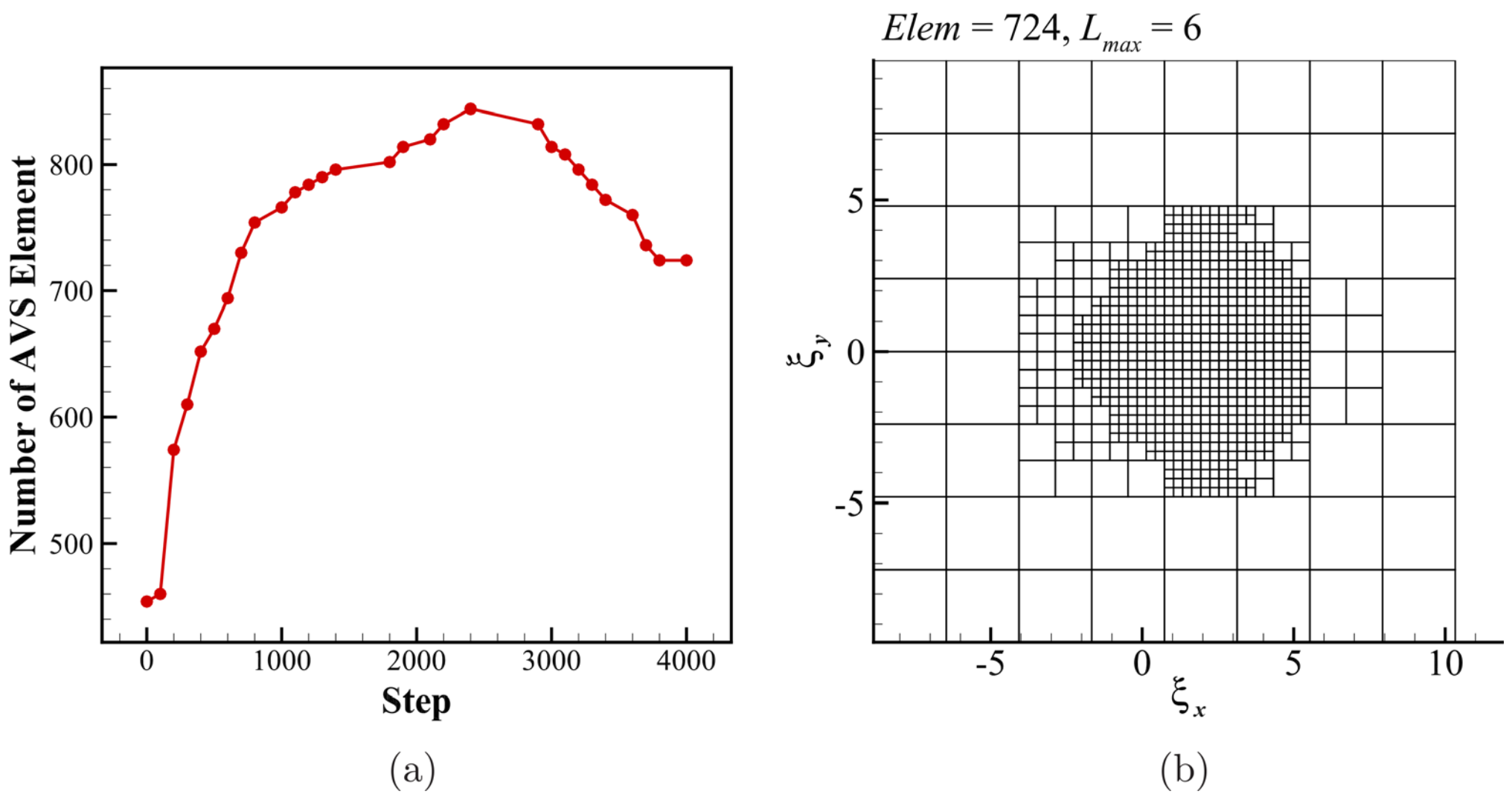

FIG. 26. AVS for the jet flow. (a) Evolution of the number of AVS elements and (b) the final AVS.

**TABLE VIII.** Computational efficiency of the jet flow.

| ... | CVS | AVS | Ratio |
|---|---|---|---|
| Elements | 10 201 | 844(peak), 724(final), 758(average) | 12.1, 14.1, 13.5 |
| Steps | 4 000 | 4 000 | 1.0 |
| Time (s) | 220 586.6 | 17 048.8 | 13.0 |

The parameter $C_1$ is 0.64 according to Eq. (34). Figure 11(a) illustrates the evolution of the number of AVS elements, and the final AVS with 586 elements is shown in Fig. 11(b). Additionally, a CVS with 7921 elements (89 × 89) is adopted to simulate the same flow. The computation time is compared in Table IV, indicating that AVS is ten times more computationally efficient than CVS. Figures 12 presents the density, pressure, temperature, and horizontal velocity (U) contours around the blunt wedge. It is evident that the results from AVS and CVS agree well with each other. The pressure coefficient, skin friction coefficient, and heat transfer coefficient along the surface of the blunt wedge are depicted in Figs. 13–15, respectively. As observed, the results from AVS and CVS are consistent and agree well with those from DS2V.

### C. Hypersonic flow over a truncated flat plate

To investigate the performance of the proposed method in handling diatomic gases, the simulation of the hypersonic flow over a truncated flat plate is conducted, with results compared to experimental data. According to the experiment of Allegre *et al.*,[42] the freestream is nitrogen gas with a Mach number of 20.2 and a temperature of 13.32 K. The density and pressure of freestream are $1.727 \times 10^{-5}$ kg/m and 0.0683 Pa, respectively. The flat plate is 100 mm long and 5 mm thick, with a wall temperature of 290 K. The Kn number based on the length of the plate is 0.0169, and the VHS model with $\omega = 0.74$ is employed.

The splitting threshold $C_1$ according to Eq. (34) is 0.17. Figure 16 displays the unstructured physical mesh, comprising 8055 cells. Additionally, Figs. 17 and 18 depict the AVS for AOA = 0° and 10°, respectively. Furthermore, the CVS with 10 201 (101 × 101) elements is also utilized to simulate the flow. The computational efficiency is shown in Tables V and VI. In comparison to the CVS, the AVS demonstrates approximately an order of magnitude faster computational performance. The pressure coefficient and heat transfer coefficient along the lower surface of the flat plate at AOA = 0° and 10° are shown in Figs. 19 and 20, respectively. The present results are in good agreement with the results of DSMC and experimental data. Additionally, the results from both AVS and CVS coincide perfectly. Furthermore, Fig. 21 presents the pressure, horizontal velocity (U), translational temperature, and equilibrium temperature contours around the truncated flat plate at AOA = 10°. It is evident that the results obtained from AVS and CVS are in excellent agreement with each other.

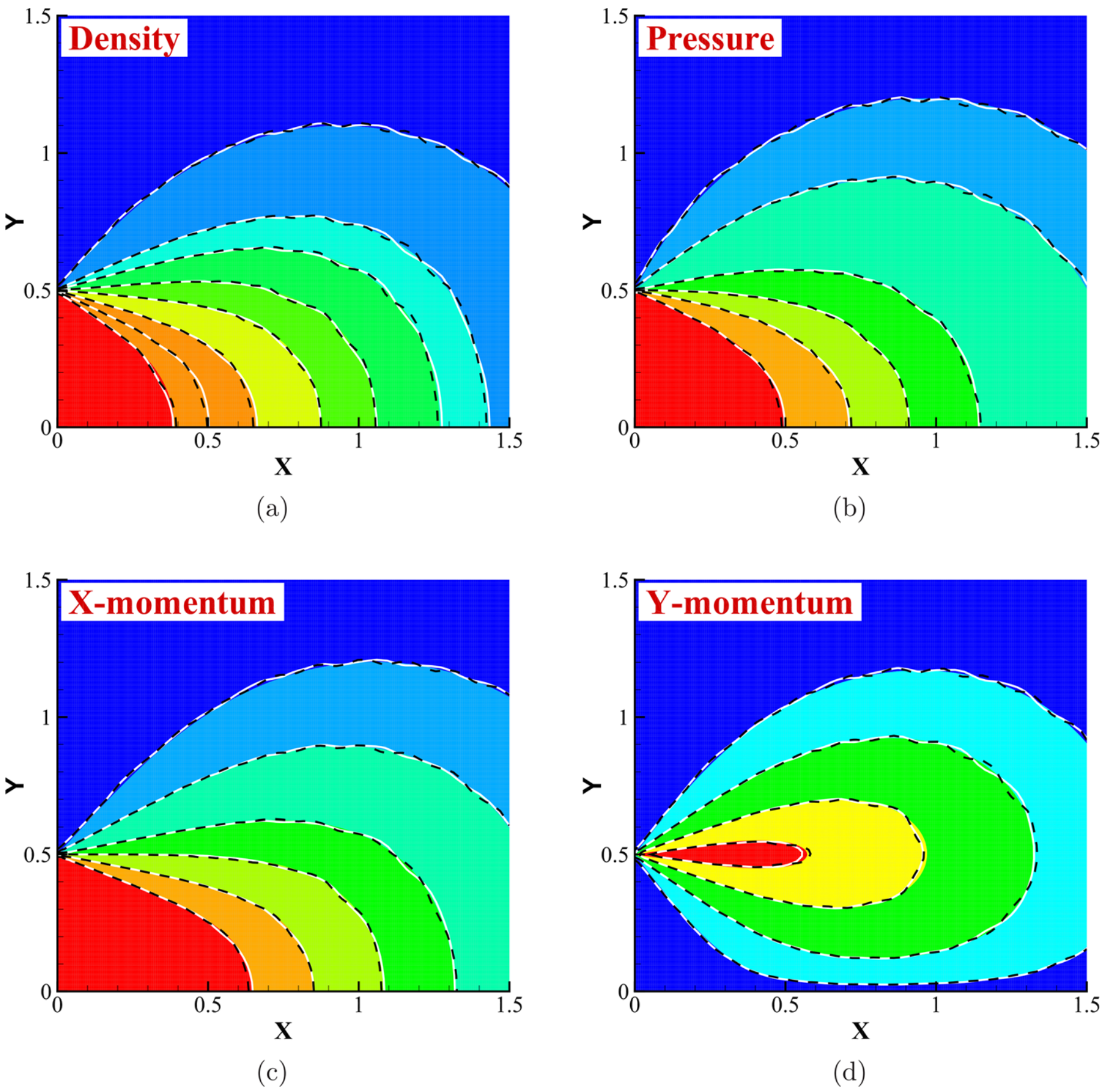

**FIG. 27.** Contours of the jet flow at $t = 0.5$ (background: analytical solutions of the collisionless Boltzmann equation; white solid lines: CVS; and black long dashed line: AVS). (a) Density, (b) pressure, (c) X-momentum, and (d) Y-momentum.

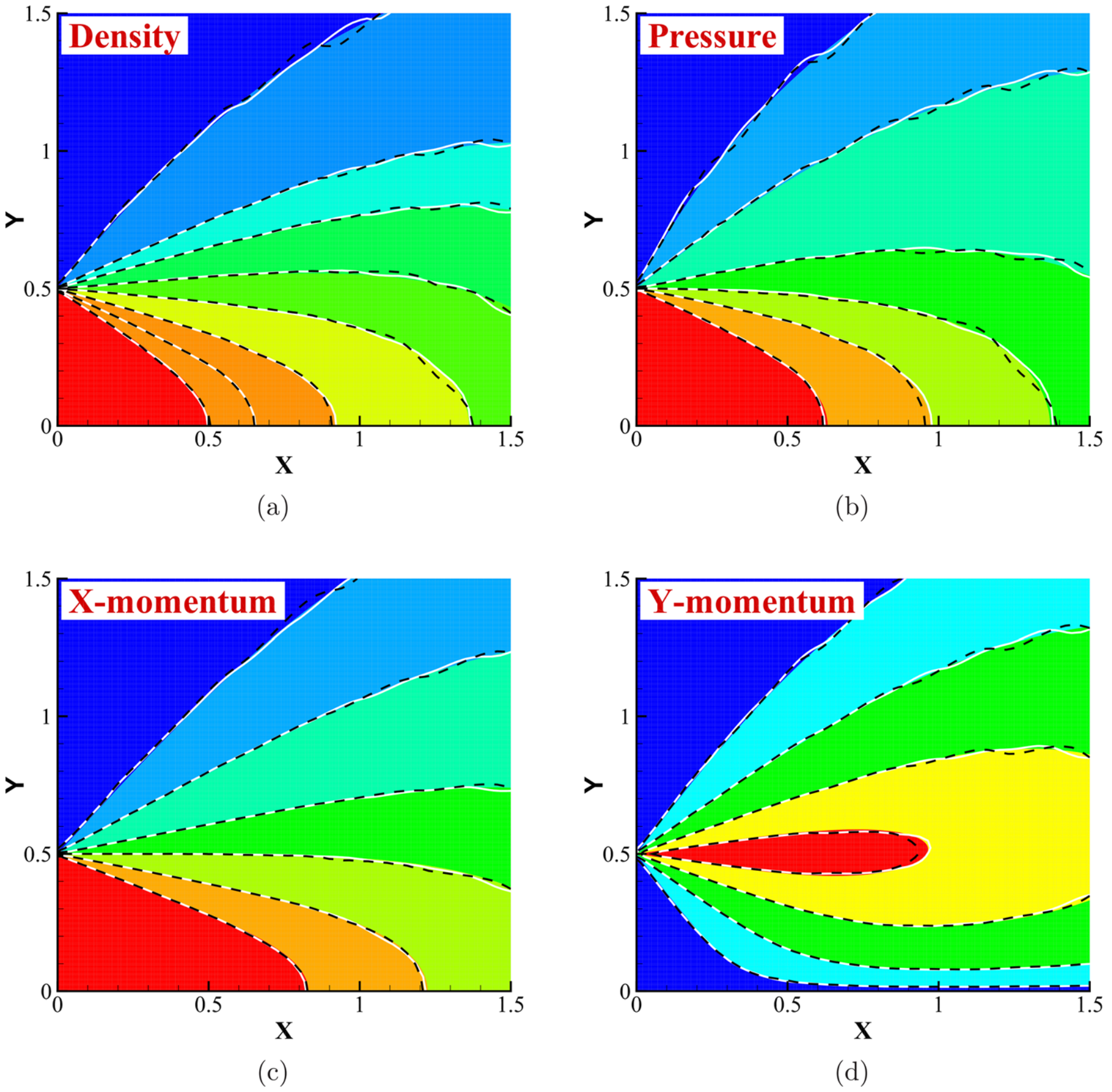

**FIG. 28.** Contours of the jet flow at $t = 1.0$ (background: analytical solutions of the collisionless Boltzmann equation; white solid lines: CVS; and black long dashed line: AVS). (a) Density, (b) pressure, (c) X-momentum, and (d) Y-momentum.

## D. Two-dimensional Riemann problem

We now test the unsteady property of the proposed method with the 2 D Riemann problem with constant initial data in each quadrant. Here, we choose the same configuration as literature,[10] where the initial condition is given by

$$(\rho, u, v, p) = \begin{cases} (\rho_1, u_1, v_1, p_1) = (0.5313,\ 0,\ 0,\ 0.4), & x > 0,\ y > 0, \\ (\rho_2, u_2, v_2, p_2) = (1,\ 0.7276,\ 0,\ 1), & x \le 0,\ y > 0, \\ (\rho_3, u_3, v_3, p_3) = (0.8,\ 0,\ 0,\ 1), & x \le 0,\ y \le 0, \\ (\rho_4, u_4, v_4, p_4) = (1,\ 0,\ 0.7276,\ 1), & x > 0,\ y \le 0. \end{cases} \quad (35)$$

Since the flow is in the free molecular regime with $\mu_0 = 10$, a CVS with $101 \times 101$ mesh points in $[-10, 10] \times [-10, 10]$ is employed to capture the nonequilibrium effects. Furthermore, a uniform mesh with $60 \times 60$ cells is used in the physical space.

Unlike the previous simulations involving nonequilibrium external flows, the Mach number varies in the 2 D Riemann problem. Consequently, determining the radius $R_{dv}$ of AVS based on the total temperature becomes impractical in this scenario. Therefore, we employ method 2 to calculate the $L_{max}$. In this case, the minimum element size is $h_{min} = 0.3$, and the approximate radius is $R_{dv,apr} = 9\sqrt{RT_1}$, where $T_1$ is the initial temperature in the first quadrant. Consequently, the corresponding $L_{max}$ and $R_{dv}$ are 6 and 9.6 $(11\sqrt{RT_1})$, respectively. Furthermore, the splitting threshold $C_1$ is set to 0.001 for this case. It should be noted that Eq. (34) is applicable only when method 1 is employed to calculate the maximum level $L_{max}$.

**TABLE IX.** The drag coefficients of sphere.

| Re | Present | Exp (air) | Relative error(%) |
|---|---|---|---|
| 19.0 | 2.04 | 2.12 | −3.77 |
| 53.0 | 1.65 | 1.69 | −2.37 |
| 80.5 | 1.52 | 1.53 | 0.65 |
| 150.0 | 1.38 | 1.37 | 0.73 |
| 210.0 | 1.33 | 1.35 | −1.48 |

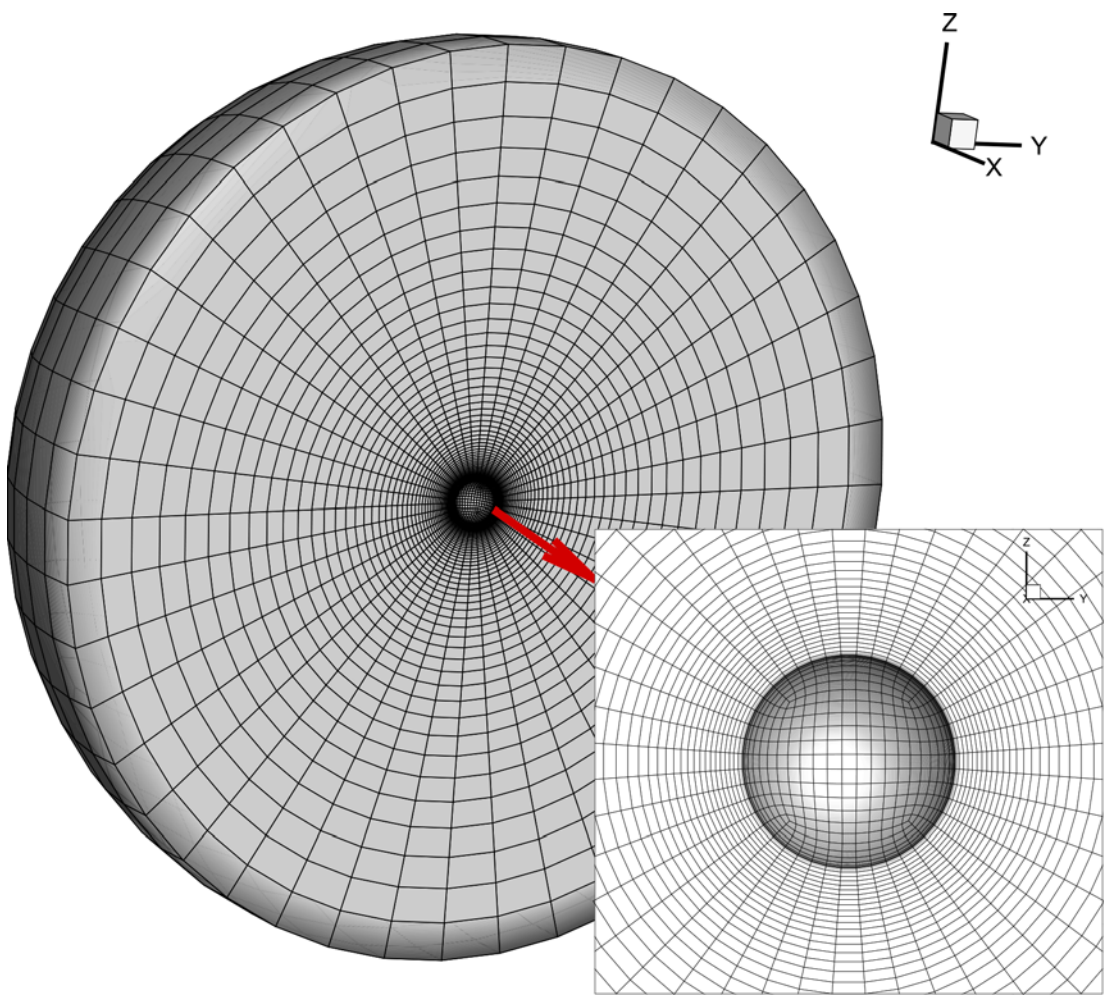

**FIG. 29.** Physical mesh for the supersonic flow over a sphere.

**TABLE X.** Computational efficiency of the sphere simulation (Ma = 4.25, Re = 210.0). The simulations with SPARTA and UVS are conducted on 2 nodes, while the simulation with AVS is conducted on a single node. The SPARTA simulation incurs significant computation time in the near-continuum regime (Kn = 0.031) due to the large number of simulated particles (114 637 232).

| ... | SPARTA (S) | UVS (U) | AVS (A) | Ratio (S/A) | Ratio (U/A) |
|---|---|---|---|---|---|
| Elements | ... | 22 670 | 3620 | ... | 6.3 |
| Cores | 128 | 128 | 20 | 6.4 | 6.4 |
| Elements per core | ... | 177.1 | 181 | ... | 1.0 |
| Steps | 80 000 | 180 | 180 | 444.4 | 1.0 |
| Time (s) | 44 281 | 3 758.4 | 2886.8 | 15.3 | 1.3 |
| Core-hours | 1 574.4 | 133.6 | 16.0 | 98.4 | 8.4 |

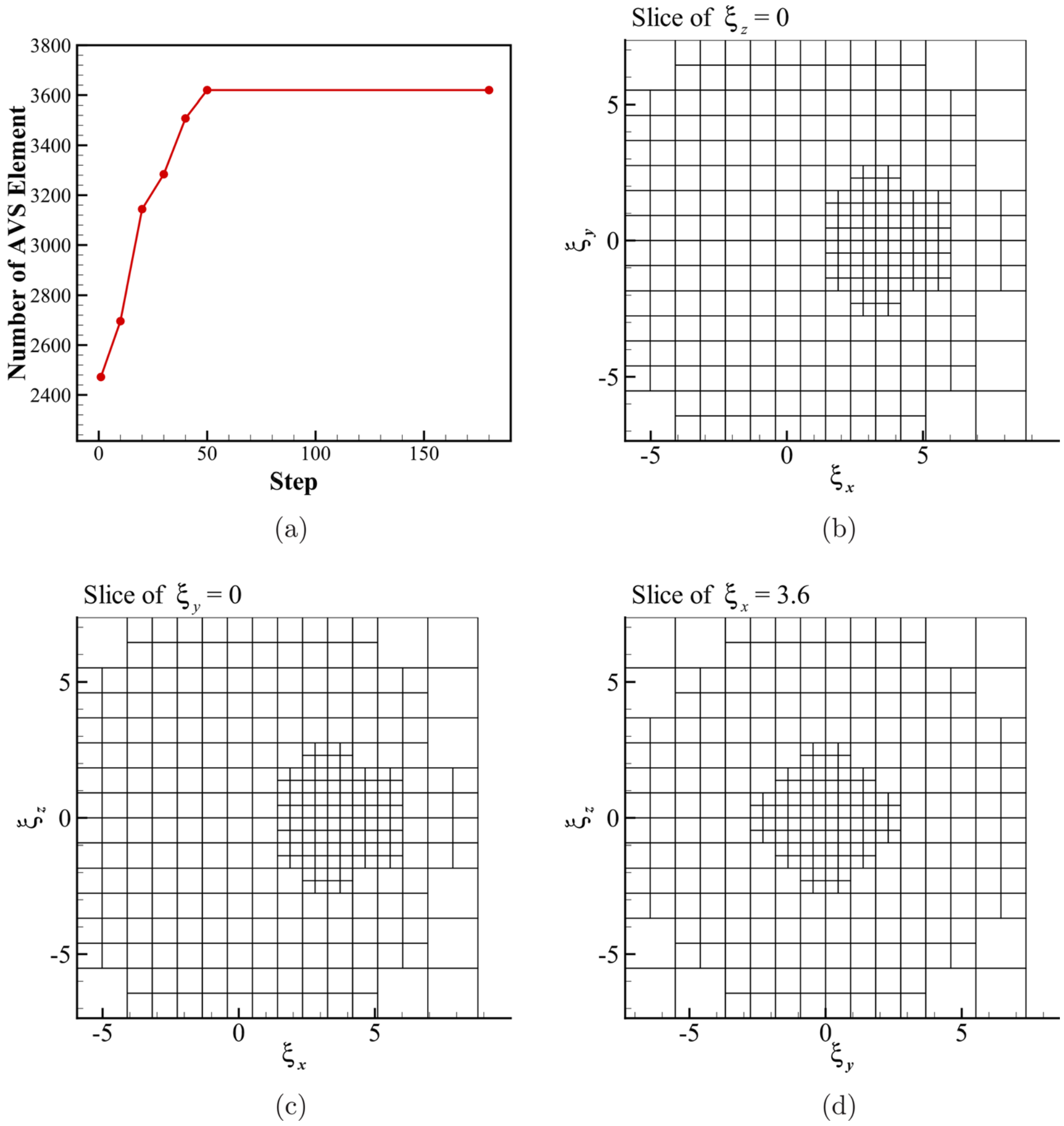

**FIG. 30.** AVS for the sphere simulation at Ma = 4.25 and Re = 210.0. (a) Evolution of the number of AVS elements, (b) slice of $\xi_z = 0$, (c) slice of $\xi_y = 0$, and (d) slice of $\xi_x = 3.6$.

Figure 22(a) displays the evolution of the number of AVS elements, and the final AVS with 673 elements is shown in Fig. 22(b). Figure 23 presents the density and temperature contours at $t = 0.15$. It is evident that the results obtained from AVS and CVS are in excellent agreement with each other. The computational efficiency is detailed in Table VII, demonstrating that the proposed global AVS achieves an order of magnitude higher acceleration than CVS.

### E. Rarefied jet flow expanding into a vacuum environment

The startup process of rarefied jet flow expanding into a vacuum environment, a practical flow topic in many domains,[43,44] is simulated to further investigate the performance of global AVS in unsteady non-equilibrium flow. The geometry of the jet flow is illustrated in Fig. 24. The width of the nozzle outlet (denoted by AB in Fig. 24), which is also the inlet of the jet flow, is $L = 1$ m. The coordinate origin (0,0) is set at the central point of the nozzle outlet, which is denoted by "O" in Fig. 24. The working gas is argon, and the hard-sphere model with $\omega = 0.5$ is adopted. The Mach number and temperature of the jet flow are 2.19 and 273 K, respectively. The physical mesh, consisting of 27 648 cells, is displayed in Fig. 25.

Since the jet flow accelerates as the gas expands into the low density environment, resulting in a changing Mach number, method 2 is employed to calculate the $L_{max}$: the minimum element size is $h_{min} = 0.3$ and the approximate radius is $R_{dv,apr} = 6\sqrt{RT_0}$. Consequently, the corresponding $L_{max}$ and $R_{dv}$ are 6 and 9.6 ($8.4\sqrt{RT_0}$), respectively.

Furthermore, the splitting threshold $C_1$ is 0.001 for this case. Figure 26(a) displays the evolution of the number of AVS elements, and the final AVS with 724 elements is shown in Fig. 26(b). It can be observed that the AVS is updated more frequently than in the previous simulations, with the number of elements peaking at 844 in step 2400, while the final AVS at step 4000 only has 724 elements. In this case, the average number of elements in AVS is 758. Additionally, the CVS with 10 201 ($101 \times 101$) elements is also utilized to simulate the flow. The computation time is compared in Table VIII, highlighting that AVS is over 10 times more computationally efficient than CVS. Figures 27 and 28 present the contours of the density, pressure, and momentum components in the X- and Y-directions at $t = 0.5$ and 1.0, respectively. It is evident that the results obtained from AVS and CVS are in excellent agreement with the analytical solutions of the collisionless Boltzmann equation.

### F. Supersonic flow over a sphere

The performance of the proposed method in 3 D flow is investigated by the simulation of supersonic flow over a sphere. Consistent with Wendt's work,[45] the freestream maintains a Mach number of 4.25 and a total temperature of 300 K, while the temperatures of the freestream and wall are 65.04 and 302 K, respectively. The working gas is nitrogen and the Sutherland formula[46] is used to calculate the viscosity,

$$\mu = \mu_{ref}\left(\frac{T}{T_{ref}}\right)^{3/2}\frac{T_{ref}+C_s}{T+C_s}, \tag{36}$$

where $C_s$ is 111 K for nitrogen and 124 K for air, and $\mu_{ref}$ is the reference viscosity at the reference temperature $T_{ref}$. The Reynolds numbers based on the diameter (2 mm) of sphere are listed in Table IX. Figure 29 shows the physical mesh consisting of 42 900 cells. The height of the first layer of the mesh is 0.01.

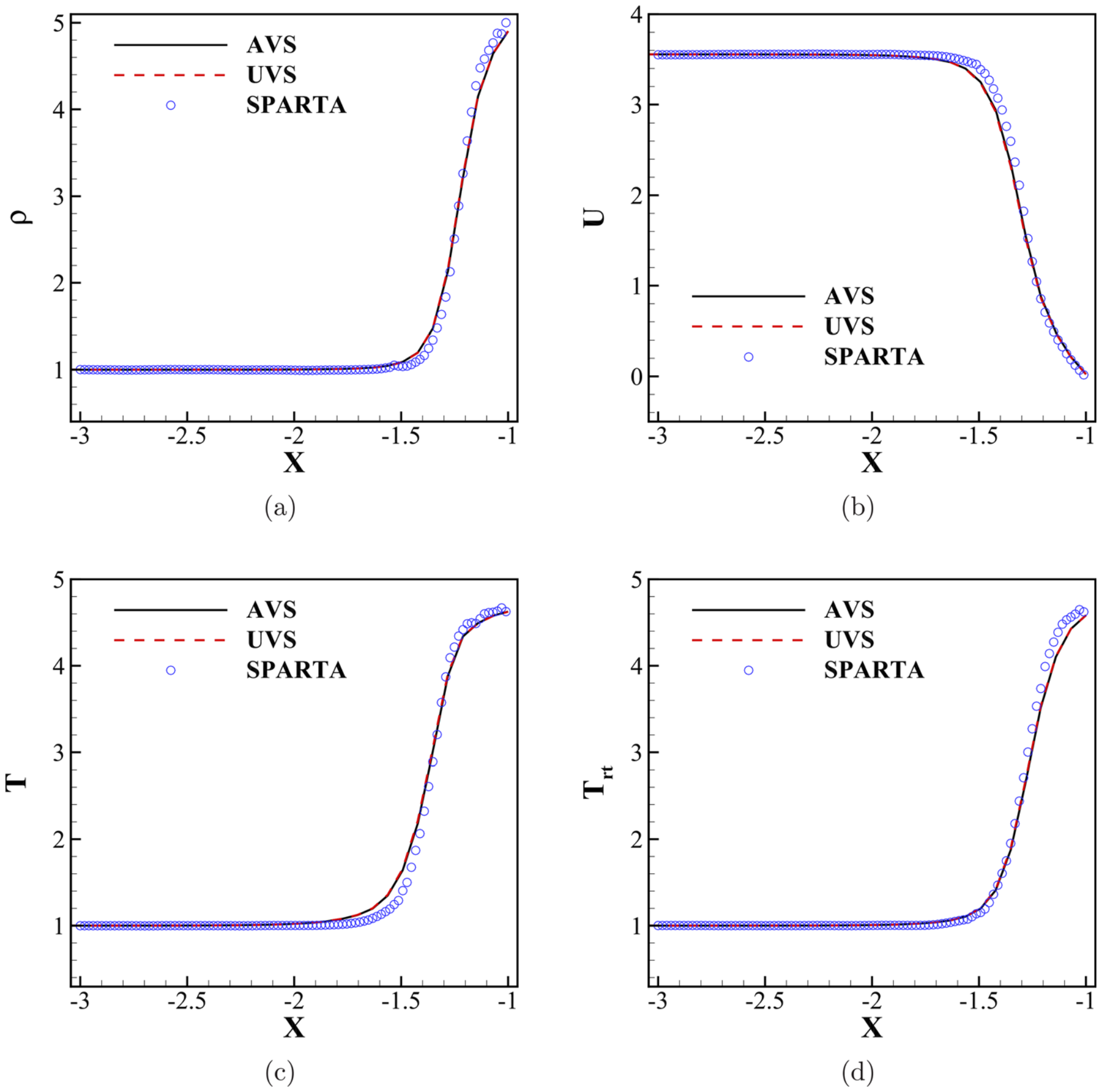

FIG. 31. Physical variables along the stagnation line in front of the sphere at Ma = 4.25 and Re = 210.0. (a) Density, (b) horizontal velocity, (c) equilibrium temperature, and (d) rotational temperature.

As the wall temperature matches the total temperature of the freestream, the stationary zone (zone B in Fig. 3), with a radius of $R_B = 3\sqrt{RT_w}$ and centered at zero, necessitates a larger domain to encompass this broader refined zone. Therefore, we employ the flexible method 2 to calculate the $L_{max}$: the minimum element size and the approximate radius are $h_{min} = 0.46$ and $R_{dv,apr} = 4\sqrt{RT_0}$, resulting in the corresponding values of $L_{max}$ and $R_{dv}$ as 5 and 7.36 ($4.8\sqrt{RT_0}$), respectively. Furthermore, the splitting threshold $C_1$ is set to 0.0001 for this case.

Figure 30 shows the evolution of the number of AVS elements and the final AVS for $Re = 210$, which contain 3620 elements. In the previous study,[41] Jiang *et al.* employed a CVS ($41 \times 41 \times 41$) with 68 921 elements to conduct the same simulation, which is approximately 19 times as many elements as the present AVS. In light of the high computational cost associated with CVS, an UVS with 22 670 elements is utilized to assess the computational efficiency of AVS. Due to the limitation of computer memory, 2 nodes (each node has a memory of 256 G and 64 cores; Central Processing Unit (CPU): AMD EPYC 7452 @ 2.35 GHz) are used when the UVS is employed. The more cores that are employed, the fewer elements each core needs to calculate. Consequently, the ratio of Message Passing Interface (MPI) communication time to total computation time increases, resulting in a lower parallel efficiency. Considering that the average number of elements of UVS per core is 177.1, 20 cores are used when the AVS is employed, so that the average number of elements of AVS per core is almost the same (181). Furthermore, to compare the computational cost with the DSMC method, the open-source software SPARTA was also used on the same compute cluster. Table X displays the computation time, revealing that AVS is ~15.3 times faster than SPARTA and 1.3 times faster than UVS. The core-hours cost of SPARTA and UVS are 98.4 and 8.4 times higher than that of AVS. Evidently, the present AVS requires less computation time and memory. The simulations using SPARTA and implicit UGKS with UVS are conducted on 2 nodes, facing challenges due to MPI communication between nodes, which can lead to longer computation times. Additionally, parallel efficiency generally decreases as core numbers increase. Finally, the flow is in the near-continuum regime ($Kn = 0.031$), resulting in a total particle number of 114 637 232 in the SPARTA simulation, contributing to the expensive computation time.

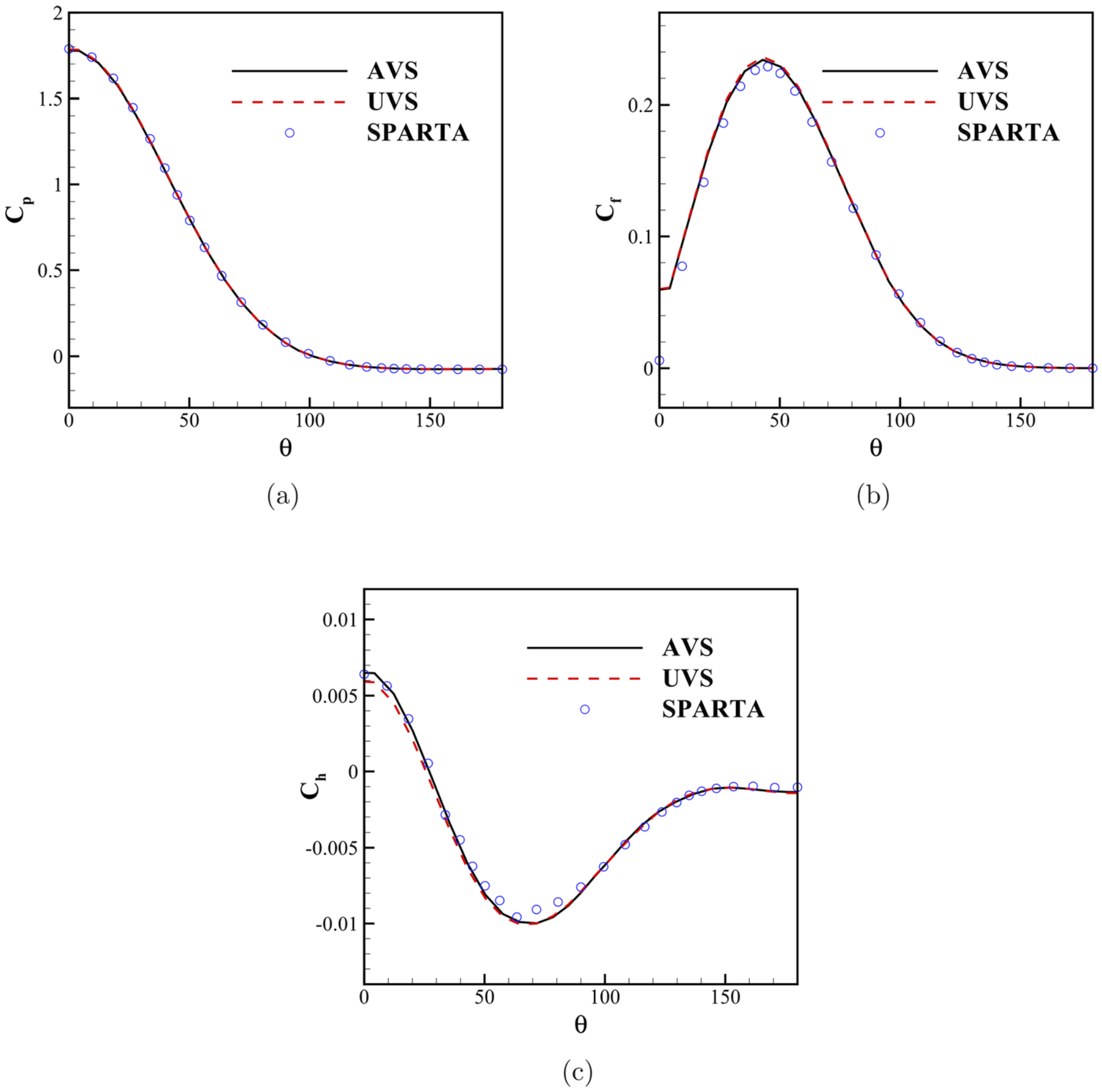

**FIG. 32.** Physical variables on the sphere surface (z = 0) at Ma = 4.25 and Re = 210.0. (a) Pressure coefficient, (b) skin friction coefficient, and (c) heat transfer coefficient.

Figure 31 shows the physical variables along the stagnation line in front of the sphere, and Fig. 32 shows the pressure coefficient, shear stress coefficient, and heat transfer coefficient along the sphere surface (z = 0). Evidently, the present results are in good agreement with those of SPARTA. Furthermore, the comparisons of drag coefficients in a wide range of Reynolds numbers are shown in Table IX, and the maximum relative error is 3.77%, indicating that the proposed method can accurately capture the behavior of 3 D multi-scale flows.

## IV. CONCLUSION

In this paper, a global AVS for general DVM frameworks is proposed. The direct mapping method is employed in the reconstruction of the distribution function in a new AVS, and the conservation is maintained by the discrete deviation of equilibrium distribution functions, in consideration of both efficiency and robustness. Compared to local AVS, the global AVS maintains a high level of parallelism and is easily extendable to implicit algorithms. Furthermore, the investigation into setting adaptive parameters and sub-zones enhances the automation of the method. Finally, a series of simulations, encompassing 2 D and 3 D, steady and unsteady, monatomic and diatomic gas multi-scale flows, have been conducted to validate the performance of the global AVS. The comparison of computation time indicates that the global AVS could improve computational efficiency by an order of magnitude and reduce memory consumption by an order of magnitude compared with the CVS.

## ACKNOWLEDGMENTS

The authors thank Professor Kun Xu at the Hong Kong University of Science and Technology and Professor Zhaoli Guo at the Huazhong University of Science and Technology for discussions on the UGKS, the DUGKS, and multi-scale flow simulations. The authors thank Dr. Dingwu Jiang at the China Aerodynamics Research and Development Center and Dr. Songze Chen at the Shenzhen Shifeng Technology Co., Ltd. for discussions on the adaptive methods. Jianfeng Chen thanks Dr. Hao Jin at the Northwestern Polytechnical University for providing the DSMC results simulated by SPARTA software. This work was supported by the National Natural Science Foundation of China (Grant Nos. 12172301 and 12072283) and the 111 Project of China (Grant No. B17037). This work was supported by the high performance computing power and technical support provided by the Xi'an Future Artificial Intelligence Computing Center.

## AUTHOR DECLARATIONS

### Conflict of Interest

The authors have no conflicts to disclose.

### Author Contributions

**Jianfeng Chen:** Conceptualization (equal); Formal analysis (equal); Investigation (equal); Methodology (equal); Software (lead); Validation (equal); Visualization (equal); Writing – original draft (lead). **Sha Liu:** Conceptualization (equal); Formal analysis (equal); Funding acquisition (equal); Methodology (equal); Project administration (equal); Resources (equal); Supervision (equal); Validation (equal); Writing – review & editing (lead). **Rui Zhang:** Software (supporting); Validation (equal); Writing – review & editing (equal). **Congshan Zhuo:** Funding acquisition (equal); Validation (equal); Writing – review & editing (equal). **Yanguang Yang:** Validation (equal); Writing – review & editing (equal). **Chengwen Zhong:** Funding acquisition (equal); Methodology (equal); Project administration (equal); Supervision (equal); Validation (equal); Writing – review & editing (equal).

## DATA AVAILABILITY

The data that support the findings of this study are available from the corresponding author upon reasonable request.